# Room Temperature Microsecond Coherence of Silicon Dangling Bonds

J. Möser[1], H. Popli[2], T.H. Tennahewa[2], T. Biktagirov[3], J. Behrends[4], W. Akhtar[1,5], H. Malissa[2], C. Boehme[2], W.G. Schmidt[3], U. Gerstmann[3]*, K. Lips[1,4]*

[1]EPR4Energy & Berlin Joint EPR Lab, Department Spins in Energy Materials and Quantum Information Science (ASPIN), Helmholtz-Zentrum Berlin für Materialien und Energie, Hahn-Meitner-Platz 1, 14109 Berlin, Germany.

[2]Department of Physics and Astronomy, University of Utah, Salt Lake City, Utah 84112, USA.

[3]Lehrstuhl für Theoretische Physik, Universität Paderborn, Warburger Straße 100, 33098 Paderborn, Germany.

[4]Berlin Joint EPR Lab, Fachbereich Physik, Freie Universität Berlin, Arnimallee 14, 14195 Berlin, Germany.

[5]Department of Physics, Jamia Millia Islamia, New Delhi, India – 110025

**Paramagnetic point defects in silicon provide qubits that could open up pathways towards silicon-technology based, low-cost, room-temperature (RT) quantum sensing. The silicon dangling bond (db) is a natural candidate, given its sub-nanometer localization and direct involvement in spin-dependent charge-carrier recombination, allowing for electrical spin readout. In crystalline silicon, however, rapid loss of db spin-coherence at RT due to free-electron trapping, strongly limits quantum applications. In this work, by combining density-functional theory and multifrequency (100 MHz–263 GHz) pulsed electrically detected magnetic resonance spectroscopy, we show that upon electron capture, dbs in a hydrogenated amorphous silicon matrix form metastable spin pairs in a well-defined quasi two-dimensional (2D) configuration. Although highly localized, these entangled spin pairs exhibit nearly vanishing intrinsic dipolar and exchange coupling. The formation of this magic-angle-like configuration involves a > 0.3 eV energy relaxation of a trapped electron, stabilizing the pair. This extends RT spin coherence times into the microsecond range in silicon required for a future spin-based quantum sensing technology.**

As quantum sensors provide exciting new avenues to overcome the inherent limitations of classical sensors[1], they have become widely used[2,3], with an ever-increasing number of quantum-enabled sensing applications expected to emerge in the near future. In particular, quantum sensors based on spin centers, i.e., highly localized paramagnetic point defects in condensed matter, will likely continue to play a key role due to their atomic scale, high sensitivity to their environment, and straightforward readability. Such spin centers provide high spatial resolution, high sensitivity, and a broad variety of physical observables. Most significantly, vacancy-related point defects in diamond[4] (NV), silicon carbide[5,6] ($V_{Si}$, $V_CV_{Si}$, $N_CV_{Si}$), and recently also boron nitride[7,8] ($V_B$), have been demonstrated to exhibit microsecond spin-coherence times at room temperature (RT) and optical readability, making them prime candidates for spin-based quantum-sensing applications. For these systems, sensing typically takes place by interaction of their spin states with the observables of their local environment, i.e., magnetic fields, while coherent quantum-control sequences are used to detect subtle changes of the sensor qubits' Hamiltonian upon coherent spin propagation. The required coherent control sequences[1] are typically realized by electromagnetic radiation pulses. The extraordinary long RT spin-coherence times of vacancy-related defects are possible due to specific, highly symmetric configurations in their respective host matrices, resulting in energy-level schemes that can be addressed optically through cycle transitions[4,9]. A purely *electric* readout of the spin information, however, is difficult to achieve by these means[10].

In contrast, some spin qubits in silicon have been shown to be suitable for electric spin readout[11,12,13]. This has been demonstrated, e.g., for charge-carrier recombination transitions between paramagnetic dangling-bond (db) defects and conduction electrons trapped at a phosphorous donor, ($P_{Si}^0$)[11]. A similar process was observed for conduction-band tail states in hydrogenated amorphous (a-Si:H)[14] and microcrystalline silicon (μc-Si:H)[15,16,17,18] as well as at the c-Si/SiO$_2$ interface[11,12,19], albeit this was achieved so far only for low temperatures ($T < 10$ K) and subensembles of randomly distributed point defects coming into appropriate distances. The resulting spin-pairs thus lack well-defined intrinsic spin-dependent charge-carrier transitions required for coherent spin-control. Aside from their random-pair nature, these spin-pair systems also suffer from short spin-coherence times with increasing temperature, as their lifetime is dramatically reduced due to thermal emission of shallowly trapped ($< 50$ meV) excess electrons. Stronger localization implies deeper band states and, thus, reduced thermal emission, but also stronger exchange coupling and by this, again shorter lifetimes and spin-coherence, as, e.g., observed for exciton states in organic semiconductors, where weak screening promotes localization and detectable spin-coherent is only seen at low temperatures[20]. Nonetheless, polaron pair states in polymer thin-film devices have been proposed for quantum sensing[21]. In non-molecular solids, however, the direct relation between localization and exchange coupling prevents such systems from this kind of application.

In this work, we report experimental and theoretical evidence for an intriguing exception to this relationship. We identify a well-defined intrinsic db-related spin-pair state in a-Si:H with strong, sub-nanometer localization, yet, near absence of internal exchange and dipolar interaction, resulting in microsecond-long spin relaxation times. By means of density-functional theory (DFT) calculations, we show that such a spin pair form as an intermediate state within the trapping and recombination process of an electron via a db. Prior to recombination, an electron is trapped in the vicinity of a neutral, i.e., singly occupied db$^0$ (see also Fig. 1b). The resulting metastable db$^{-*}$ complex provides an intrinsic paramagnetic structure in a quasi-two-dimensional so-called 'magic' configuration with almost vanishing dipolar coupling. This surprising result shows that spin-dependent electronic transitions involving single-point defects in a-Si:H can be utilized for sensitive electronic spin readout without shortening the spin coherence. This unique feature makes the paramagnetic db-related spin-pair state a highly suitable candidate for RT spin-based quantum sensing.

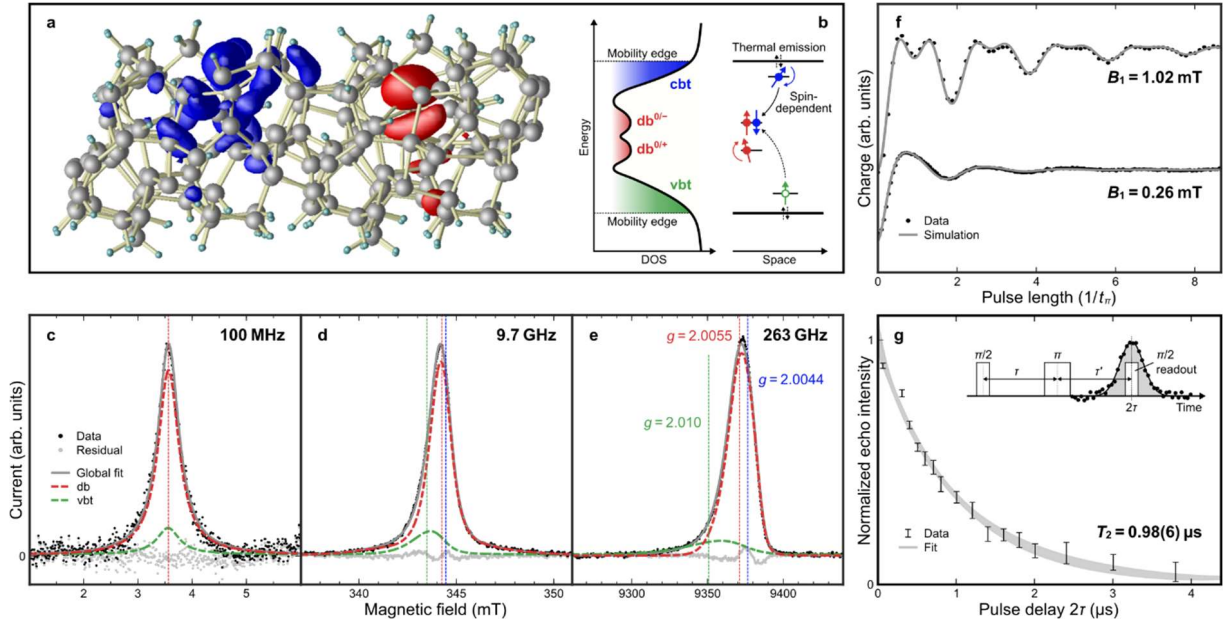

**Fig. 1. Multifrequency pulsed-EDMR spectra on an *a*-Si:H/*μc*-Si:H heterostructure.** (**a**) DFT-calculated spin-distribution of a neutral dangling bond (db$^0$, red) and a distant conduction-band tail (cbt, blue) state in an a-Si:H matrix. (**b**) Schematic sketch of the recombination process (right) and of the density of states distribution of the involved defect states, i.e., of their charge-transition levels (left). (**c**)–(**e**) Room-temperature (RT) field-swept pulse electrically detected magnetic resonance spectra of an a-Si:H/μc-Si:H heterostructure *pin* device, measured in the dark, at a forward bias of 1.1 V and at microwave-excitation frequencies of (**c**) 100 MHz, (**d**) 9.6 GHz and (**e**) 263 GHz. The red line shows a best fit comprising essentially a single db resonance (red dashed line, spin-Hamiltonian parameters from Ref. 22) and a small (12 %) contribution associated with valence-band tail (vbt) states. Dotted markers indicate the average resonances of dbs ($g = 2.0055$), cbt ($g = 2.0044$) and vbt states ($g = 2.01$) known from literature[23,24,25,26]. (**f**) Electrically detected Rabi nutations recorded at 9.6 GHz at weak (0.26 mT) and strong (1.02 mT) $B_1$ excitation fields. The appearance of an additional beating (*spin-locking*) component with increasing $B_1$ is a fingerprint of a weakly coupled *two-spin S = 1/2* pair. (**g**) Electrically detected Hahn-echo decay of the signal shown in (**c**) and (**e**), showing an exceptionally long coherence time $T_2 = 1$ μs at RT. For further details see text.

# Electrical detection of Si dangling bond spins at room temperature

In the following we present multifrequency (100 MHz–263 GHz) pulsed EDMR (pEDMR) spectra measured in the RT dark recombination current of an a-Si:H pin bipolar injections device, identical to state-of-the-art silicon pin solar-cell structures (for details of the sample and experimental setup see Refs. 27–29). Our data show that, (*i*) the electric readout current at RT exhibits surprisingly long spin-coherence times of about one µs and (*ii*) the spectra can essentially be reproduced by the characteristic electron-paramagnetic-resonance (EPR) fingerprint of a db in its neutral ground state (db$^0$)[22,30,31]. Figures 1c–e show multifrequency pEDMR spectra (100 MHz, 9.6 GHz, and 263 GHz, respectively) centered around the known *g*-value distributions[22] expected for a db state in a-Si:H. By globally fitting the spectra using the *g*-tensor and hyperfine (hf) splitting signatures of all known intrinsic EPR lines in a-Si:H, i.e., dbs as well as conduction- (cbt) and valence-band tail (vbt) states[22,25,32], we obtain a best global fit using a single db resonance and an additional, though small (12 %) contribution of a vbt resonance ($\bar{g} \approx 2.01$). A bootstrap routine proves the extreme robustness of the fit (see SI, Notes 1 and 2, for more and the related statistics). In essence, the additional contribution appears too small to explain the underlying recombination process, rendering dbs as predominant spin-pair partners within the detected recombination transitions.

Dangling bonds are ubiquitous point defects in various silicon-based solid-state phases, not only in a-Si:H, but also in thin µc-Si:H films, at surfaces[33,34] (e.g., grain boundaries) and at a-tetracene/c-Si[98] and Si:H/c-Si interfaces[26,100], where they are referred to as so-called P$_b$ centers[35,36,37]. The db$^0$ in a-Si:H is a prototypical dangling bond. The overwhelming part of the spin density is localized in a *sp$^3$*-hybridized orbital localized at a threefold coordinated Si atom, as exemplarily shown in Fig. 1a (red). The disordered a-Si:H network leads to nanoscale density fluctuations[38] of the material eventually developing localized cbt and vbt states, with dbs (in form of db$^-$, db$^0$, db$^+$) being strongly localized and having energies somewhere in the bandgap (Fig. 1b)[39]. Neutral db$^0$ provide singly occupied midgap states and thus act as traps for free electrons (db$^{0/-}$) and holes (db$^{0/+}$). It is this charge capture that allows an electric readout of the spin information by means of recombination currents. The dynamics of this process can be described by Shockley-Read-Hall (SRH) statistics[40,41], with respective capture-cross sections for db$^{0/-}$ and db$^{0/+}$ transition energies[39]. The SRH model specifically assumes that the spin-realignment time upon *direct* charge capture[40] is very fast for both electron trapping as well as structural relaxation, resulting in a rapid loss of spin coherence.

The formation of specific spin-pair configurations *prior* to the final capture process remains as an alternative for longer-lived spin systems. In this context, hypothesis on *excited* db$^-$ states have been already made in µc-Si:H films and at the SiO$_2$/c-Si interface[16,19,42] but never corroborated in these systems nor in a-Si:H. Instead a spin pair, more precisely a cbt-db pair, has been suggested[23,24,25] given by a shallow cbt state (Fig. 1a–b, blue) within nanometer proximity to a db$^0$ (red). However, due to the small trapping energy of cbt, the thermal reemission time to the conduction band at RT would be in the picosecond range[103]. The spin-pair coherence times of such pairs would, thus, again be much too short for any practical RT application.

In the present EDMR study we were able to detect the db-related magnetic response in a-Si:H with much longer spin-coherence times. Electrically detected Hahn-echo experiments[43,44] (Fig. 1g) result in slow echo decays, which reveal an exceptionally long RT coherence time of 0.98(5) µs comparable to an a-Si:H db coherence time [Fehr et al, PRL 2014]. The apparently long-lived spin state even at RT fulfills the essential prerequisites for using db-related defects in a-Si:H as spin sensors.

## Spin-pair mechanism and spin coherence

The applicability of the observed spin center as RT spin sensor depends on the exact nature of the spin-dependent recombination mechanism behind the observed RT EDMR resonance, i.e. the wavefunctions of the involved electronic states, coupling strengths, the spin dynamics as well as the dynamics of spin-dependent electronic transitions. To obtain this information, we begin by measuring the spin-coupling by electrically detected Rabi (ED-Rabi) nutations[45] (Fig. 1f) as a function of the driving microwave field ($B_1$). We clearly observe a Rabi-oscillation pattern that needs a few microseconds for decaying away, reflecting long coherence times of the associated spin pair. At low $B_1$, we observe a damped oscillation at a Rabi frequency $\Omega = \gamma B_1$ (lower trace). With increasing $B_1$, an additional beating component at $\Omega = 2\gamma B_1$ emerges (upper trace). Such characteristic beating arises under *spin-locking* conditions, when two almost uncoupled spins are excited simultaneously due to the increased excitation window[46].

A detailed analysis of the magnetic-field dependency of the ED-Rabi nutations (Fig. 2) further confirms this scenario. The respective Rabi-frequency profiles obtained from the fast Fourier transform (FFT) are particularly sensitive to Larmor-frequency distributions and spin-spin coupling strengths[46,47,48,49,50]. The observed patterns in Fig. 2a show again the required appearance of the second beating frequency that emerges at $\Omega = 2\gamma B_1$ (cf. Fig. 1f). Such patterns are in fact indicative of a *two-spin* ($s=½$) pair transition, where both spins are only *weakly coupled* by exchange ($J$) and dipolar ($D$) interaction[51]. Stronger spin-spin couplings would lead to characteristic Rabi-frequency components (e.g., $\Omega = \sqrt{2}\gamma B_1$ for strong dipolar coupling)[50,52], which are not present in Fig. 2a.

The observation of RT coherence times in the range of 1 µs implies thermal emission times in excess of $10^{-6}$ s, and thus, binding energies of the spin pair of around 0.4 eV[103]. This immediately rules out tunneling transitions of charge carriers trapped in nearby shallow cbt or vbt states (<0.2 eV below (above) the conduction (valence) band edge), as proposed by Dersch and others[23,24,25,53,54]. This is in line with the absence of any cbt/vbt contribution in the RT EPR and EDMR spectra of undoped silicon[22,23,24,25] and also in agreement with the invariance of the EPR lineshape with respect to illumination[55] and charge injection. Ruling out spin-dependent transitions involving cbt and vbt states, the only remaining explanation of the EDMR and ED Rabi-nutation signatures are weakly spin-coupled pairs of electron states with $S = 1/2$ that *both* resemble the signature of an intrinsic db⁰.

Having this in mind, we simulated the magnetic-field-dependent Rabi-nutation patterns for various spin-pair scenarios. (For simulation-details see Refs. 48, 50 and 56 and the SI, Note 3). Figs 2c and d show the simulated FFT map for a pair of two db-like spins (i.e., both described by a db⁰ $g$-tensor) with exemplarily assumed coupling strengths $D = J = 2$ MHz. The simulation nicely reproduces both the $\Omega = \gamma B_1$ and the $\Omega = 2\gamma B_1$ features observed experimentally, as seen in Fig. 2a and b, revealing excellent agreement of the relative peak intensities. Note that the latter are highly sensitive to the Larmor-frequency distributions of the paired spins. In the Supporting Information, SI, Fig. S3, we included numerical simulations of Rabi patterns for different, alternative two-spin models, showing significant deviations between experiment and simulation, giving further evidence for the db-related spin-pair model.

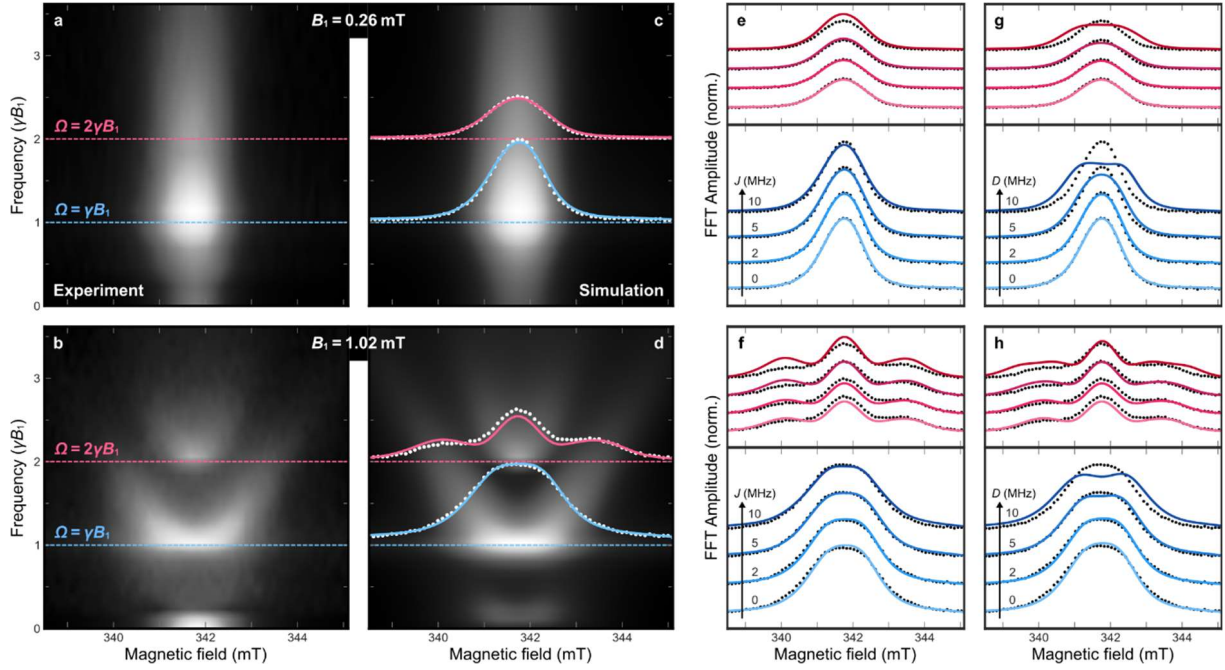

**Fig. 2. Evaluation of the spin-pair mechanism by comparison of ED-Rabi-nutation measurements to numerical simulations.** **(a)–(b)** Fast-Fourier transforms (FFT) of experimental ED-Rabi-nutation data as recorded at low (a) and high (b) values of $B_1$, as function of the applied static magnetic field. **(c)–(d)** Plot of numerically simulated FFT Rabi patterns corresponding to the experimental conditions in (a) and (b), respectively, assuming an ensemble of weakly coupled ($D=2$ MHz, $J=2$ MHz) pairs formed by two db-like spin-1/2 states (see text for details). The overlaid spectra compare the experimental (dotted) and simulated (line) cross-sections at Rabi frequencies $\Omega = \gamma B_1$ and $\Omega = 2\gamma B_1$ (marked by the blue and pink dashed lines, respectively). **(e)–(h)** Numerically simulated cross-sections $\Omega = \gamma B_1$ (blue) and $\Omega = 2\gamma B_1$ (pink) as a function of the coupling strengths $J$ (e, f, $D=0$ MHz) and $D$ (g, h, $J=0$ MHz) at low (e, g) and high $B_1$ (f, h), illustrating the effect of $J$ and $D$ on the shape of the Rabi pattern and, in particular, the evident deviation between experiment and simulation for increasing values of $J$ and $D$.

The observed Rabi frequencies and their respective peak intensities also depend strongly on the spin-spin coupling strengths $J$ and $D$. Figures 2e–h illustrate this effect for coupling strength between 0 and 10 MHz. The comparison to the experimental data immediately reveals that coupling strengths must be smaller than 5 MHz. Higher absolute values of $J$ produce increasing peak intensities at $\Omega = 2\gamma B_1$, at both low and high $B_1$ (Fig. 2e–f), whereas increasing $D$ results in broadening and splitting the $\Omega = \gamma B_1$ peak (Fig. 2g–h).[48,50]

Fig. 2d exposes an asymmetry in the field-dependent Rabi spectra, in particular apparent at high $B_1$ excitation. This asymmetry is not reproduced by the simulation (Fig. 2d, f, h) using the db$^0$ $g$-value distribution known from EPR[22]. However, several effects can alter the shape of the Rabi pattern, including slight modifications of the $g$-tensor and $g$-strain values of either one of the two paired electron state ensembles or $g$-value correlations within the spin ensembles[49,56].

Based on the results of multifrequency EDMR and ED-Rabi nutations, we conclude that the only explanation of the observed RT signal is a spin-pair mechanism based on a weakly coupled ($D, J \approx 2$ MHz) pair of two paramagnetic states with $S = 1/2$ that both exhibit an intrinsic-db resembling $g$-tensor. Spin-dependent transitions via dbs only, e.g., hopping along chains of dbs, has been observed in high-defect density a-Si:H and μc-Si:H[57,58]. In the present case, however, such pairs of distant db can be safely excluded, considering that the defect density of the specimen under study was well below $10^{15}$ cm$^{-3}$, rendering direct db-pair processes (even if assuming uniform distribution) very unlikely. In fact, local hopping in clusters and voids (db-chain hopping transport) is possible, but cannot explain the observed effects in the low macroscopic conductivity of a-Si:H[39]. Moreover, the bias dependence of the EDMR signal reported here is nearly identical to what has been reported in a-Si:H and μc-Si:H *pin* solar cells, where electron charge capture at single neutral db states is dominant[59,60,61]. We can thus exclude spin-dependent hopping or tunneling between distant db states as a conduction pathway.

Instead, we explore in the following a scenario that is based on the formation of a specific spin-pair precursor state with almost vanishing, but *non-zero* spin-spin interaction. In this weak-coupling regime, the wavefunctions are no longer eigenfunction of $S^2$ and $S_z$. Thus, the full Hamiltonian must be diagonalized in order to obtain the true eigenfunctions[51]. We evaluate this possibility by applying *constrained* DFT onto electron capture of neutral dbs in a-Si:H. The transfer of an extra electron to an already previously disorder-elongated Si–Si bond in the backbond vicinity of an atom with db yield a further elongation of the affected bond and allows the spin system to minimize the total energy. In the past, such effects were suggested to result in slow trapping procedures, but could never be verified experimentally[62,63,64]. The resulting spin state can be interpreted as a metastable db$^{-*}$ spin-pair state, built up by two db-like contributions, exhibiting a very specific and weak exchange coupling. Following this qualitative model, we obtain the following DFT-predictions for the trapping process and the properties of the involved paramagnetic states.

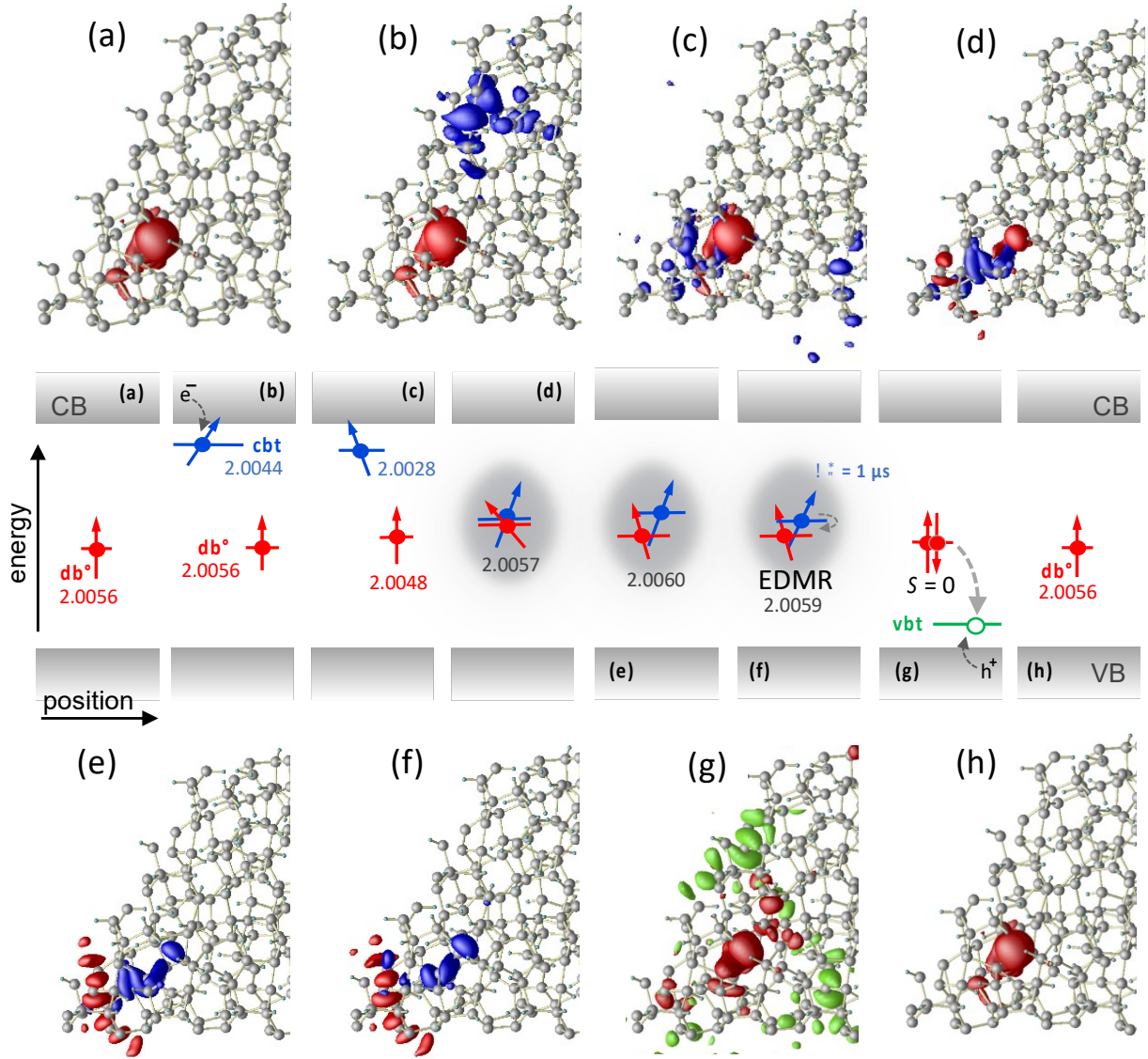

**Fig. 3. Illustration of localized electronic states involved in charge trapping and recombination of a cbt electron at a db in a-Si:H, as revealed by DFT calculations.** Starting from a neutral dangling bond db⁰ **(a)** and distant db (red) + cbt (blue) pair **(b)**, the cbt becomes mobile through thermal activation (pico- to nanosecond time scale) and **(c)** is partially trapped at the db. **(d)** Upon trapping, the db orbital is deformed, resulting in a two-spin pair of *energetically aligned* and *spatially entangled* (light-blue background shaded) spin states[65] with adapted *g*-tensors, **(e)** before the two-spin state is energetically stabilized, **(f)** and the metastable magic configuration is formed whose spin state can be probed by EDMR. **(g)** Finally, after a long (*order of* µs) lifetime, the spin pair transforms into a diamagnetic db⁻ pair with a doubly occupied db-orbital (red) giving rise to electron accumulation attractive for recombination with holes (green), **(h)** resulting again in a neutral db⁰ state. The *calculated* angular averaged *g*-values for the respective paramagnetic structures are also shown (full tensor displayed in Tab. S1 in the SI). Colors related to that of the arrows in the energy level diagrams (center row), i.e. (apart from the vbt in green) red is always the energetically lower spin, blue the higher one. **Center row:** schematic illustration of the evolution of the eigenstate energies shown in **(a)** through **(h)**. Upon trapping, **(b)**–**(f)**, the energy of the trapped electron is lowered by about 0.4 eV. In green (vbt), red (db), and blue (cbt), respectively, we denote the corresponding spin states, respectively. The DFT-calculated angular averaged *g* values for the spin states are also given. The *g* values for the entangled states (**d, e, f**) belong to the entire two-spin pair (individual *g* tensors are no longer well defined) and compare well with the experimentally determined value of 2.0057.

## Density-functional theory (DFT) analysis

Figure 3 illustrates the cascade of electronic transitions in the course of a cbt electron trapping at a neutral db$^0$ in a-Si:H (see Ref. 26 and Methods section) as well as the subsequent recombination with a vbt hole (Fig. 3g) which is not spin-dependent. In principle, it is also possible that first a vbt hole is captured at a db$^0$ state, but this process is experimentally shown to be less probable and is believed to lead to the small 12% contribution of a vbt/db-like state shown in Figs. 1c-e.

The illustrations are based on DFT calculations, which have also revealed predictions for the *g*-tensors (see Fig. 3, center row for the angular averaged values, and Tab. S1 for the full tensor) as well as the energy level of the displayed electronic states. Upon trapping at a db site, both the cbt electron (blue) but also the initial db (red) change in localization and shape, while this spatial encounter is accompanied by a stepwise alignment of the energy levels (Fig. 3, center row). Notably, in an intermediate state (Fig. 3d) the two spin states become energetically almost degenerate and spatially entangled (non-separable), explaining the experimentally observed *g*-tensor adaptation. The latter is thus a direct consequence of the trapping procedure and the related structural relaxation.

This qualitative result is independent on the lateral size of the simulation box (i.e., the supercell) and appears already in minimum-size supercells of about 2 nm × 2 nm size. Thereby, the energy lowering of the cbt state plays a central role. Distant (quasi free) cbt electrons are found rather close to the conduction-band edge (Fig. 3b). The ensemble of such shallow cbt states forms a defect band which follows the conduction-band minimum. Its dispersion of about 0.1 to 0.2 eV reflects both the statistical variations of the cbt states and the delocalized nature of their individual wavefunctions (Fig. 3c, blue). Upon trapping, i.e., the localization in the direct neighborhood to a db state, the cbt state is stepwise lowered in energy by up to 0.5 eV, but the corresponding state still retain a considerable dispersion. Both effects are facilitating entanglement, i.e. coherent coupling of the two involved spin states. Assuming thermal effects to be predominant, the maximum size of the energy lowering (0.3–0.5 eV) is fully consistent with the observed increase of spin-decoherence lifetime in silicon at RT by several orders of magnitude into the microsecond regime. The resulting structural relaxation explains a robust coherent coupling, while, at the same time, providing small dipolar exchange and spin coupling. According to our total energy calculations, the $J=J_{dip} + J_{ex}$ (dipolar and exchange, respectively) parameter is very small and hovers around zero, with its sign depending on the particular structural relaxation details within the a-Si:H network, followed by a delicate coupling of the energetically nearly degenerate spin states. The apparently dominating, but small dipolar exchange coupling $J_{dip}$ is a direct consequence of the spatial distribution of the trapped electron's spin and the modified db spin providing a specific quasi-planar geometric arrangement of differently oriented spin distributions. The same is true for the spin-spin part $D$ of the zero-field splitting. In an intuitive geometric picture, the resulting $J_{dip}$ and $D$ parameters can both be rationalized by approximating the two contributing spatially extended spin distributions by magnetic dipoles (spin vectors $S_i$) at point positions $r_i$, oriented along their spatial extension (see Fig. 4). The energy of the dipolar coupling (the dipole-dipole interaction) can be thus written as[51,70,71]

$$J_{\text{dip}} = \frac{\lambda}{|\vec{r}_{21}|^3}\left[\left(\vec{S}_1\cdot\vec{S}_2\right) - 3\left(\vec{S}_1\cdot\frac{\vec{r}_{21}}{|\vec{r}_{21}|}\right)\left(\vec{S}_2\cdot\frac{\vec{r}_{21}}{|\vec{r}_{21}|}\right)\right] \quad \text{(eq 1)}$$

$$= \frac{\lambda}{|\vec{r}_{21}|^3}|\vec{S}_1||\vec{S}_2|\cdot\left[\cos\alpha - 3\cos\beta_1\cdot\cos\beta_2\right]$$

where $\lambda$ is a specific coupling constant, $\alpha = \angle(S_1, S_2)$ denotes the angle between the two spins, and $\beta_i = \angle(S_i, r_{21})$ describes their orientations with respect to the relative distance vector $r_{21} =$ $r_2 - r_1$ of the pair partners. Magic angles as conditions under which the anisotropy-induced coupling vanishes are well known in different context[65,66,67,68]. Usually, this refers to the situation where two parallel spins are assumed ($\alpha = 0°$) and the angle denoted their common orientation $\beta$ ($=\beta_1=\beta_2$) with respect to the distance vector $r_{21}$ satisfies the condition $1 - 3\cos^2\beta = 0$, e.g., and, thus, $\beta = 54.74°$.[69] There are, of course, plenty of other 3D configurations with vanishing dipolar coupling[70,71]. In fact, the three angles only need to comply with the condition

$$\cos\alpha - 3\cos\beta_1 \cdot \cos\beta_2 = 0. \text{ (eq 2)}$$

Additional assumptions, e.g., $\alpha = \beta_2$, yields *classes* of 'magic' configurations, i.e., sets of situations with small or vanishing *J (and D)*. In fact, a closer analysis of the spin distribution of the spin pair in Fig. 3f shows a specific topological, quasi-planar (2D) configuration where the general *magic angles* condition is fulfilled for $\alpha = \beta_2$, i.e.,

$$1 - 3\cos\beta_1 = 0, \text{ (eq. 3)}$$

when $\beta_1 = 70.8°$ and $\alpha = \beta_2 = \beta_1/2 = 35.4°$ (see Fig. 4c,c').

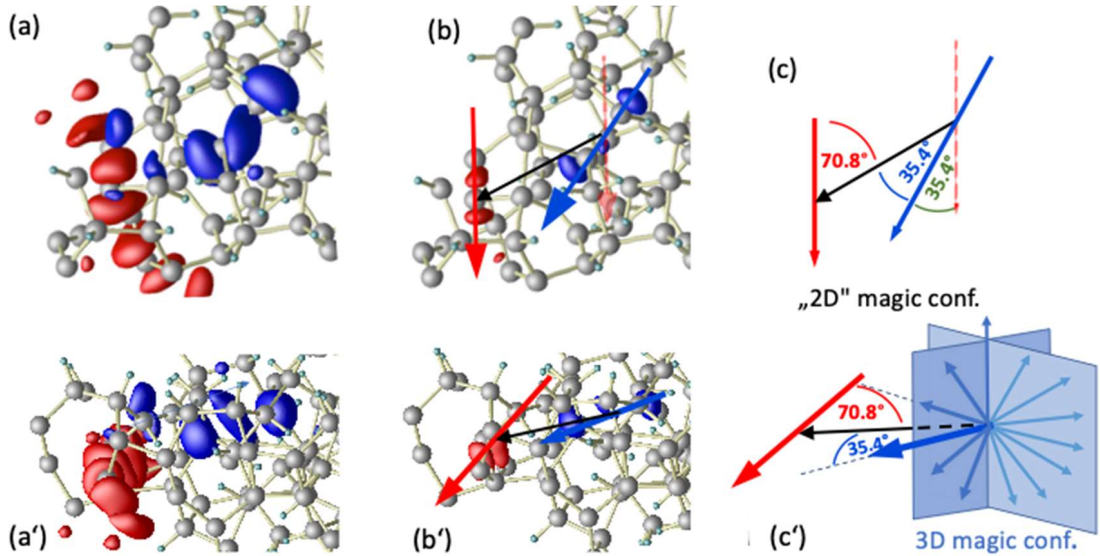

**Fig. 4. Weakly coupled spin pair in a-Si:H in quasi-2D magic configuration.** Top view (**a – c**) and side view (**a' – c'**) of the spin-distribution of the two involved orbitals (blue, red). Reduction to the spin-densest regions (**b, b'**) allows to identify the two spin-vectors $S_1$, $S_2$ (center and average orientation), fulfilling the condition for a quasi-2D magic configuration, where the dipole-dipole interaction is perfectly cancelled out (**c, c'**). Note that beside the quasi-planar (2D) magic configuration, there are many other, non-planar (3D) configurations, where the second spin $S_2$ lies in one of the two orthogonal planes indicated in blue (**c'**). Here, the condition $1-3\cos\beta_1 = 0$, i.e., $\beta_1 = \angle (S_1, r_{21}) = 70.8°$ is necessary for vanishing dipole-dipole interaction (**c'**). The quasi-*planar* (2D) magic configuration also requires $\beta_2 = \angle (S_2, r_{21}) = \beta_1/2 = 35.4°$(**c**).

At first glance, the arrangement of two spins in this quasi-2D magic configuration appears to be rather fortuitous. We have modelled the db$^{-*}$ in different supercells and at different dbs within varying local strain and lattice order and obtained identical qualitative results, resulting either in a diamagnetic db$^-$ or in a spin pair within quasi-2D magic configuration. Given that a certain degree of disorder (and strain) will exist in the vicinity of the atom with db, a spin pair with almost vanishing $J$ and $D$ couplings is always formed prior to the diamagnetic db$^-$ state. Obviously, the driving force behind this very specific coupling is the steric, energy-lowering interaction between the neutral db and the trapped electron. The extra electron profits from the flexibility of the disordered a-Si:H network, and is transferred to one of the backbond Si atoms resulting in characteristic Si---Si-db complexes, with strain-reducing elongated Si-Si bond. Thereby, the charge of the two unpaired electrons is arranged in such a way that Coulomb as well as exchange energy, are minimized. As a result, the wavefunctions of the db and the trapped cbt appear to be almost orthogonal, but non-separable in a quantum-mechanical sense (i.e., in Hilbert space). In real space, this is reflected by the entangled magic configuration (see Fig. 4), where the specific relative arrangement of the two spins results in almost vanishing spin-spin and exchange interactions. From an EPR point of view, they can, therefore, be described by a pair of weakly coupled, localized paramagnetic states[51], each with spin $S=1/2$.

## Conclusions and Outlook

We present multi-frequency pEDMR experiments that probe different RT db recombination hypotheses in a-Si:H under bipolar injection conditions. We have scrutinized these different spin-dependent charge-carrier recombination pathways by comparison of the EPR parameters ($g$-tensors and ZFS) as well as coherent spin-motion signatures, i.e., observations of electrically detected, resonantly driven spin-Rabi oscillations, with predictions from DFT for the same observables. It is shown that electron trapping at neutral dbs results in db$^{-*}$ spin pairs, where the extra electron is transferred to one of the backbond Si atoms forming characteristic Si–Si–db complexes, thereby almost entirely annihilating dipolar and exchange spin coupling. Notably, this specific recombination precursor state in a specific magic-angle like configuration can always be formed, despite and even because of the disorder that exists in the backbond-vicinities of the db atoms. Such entangled configurations are thus expected for various silicon morphologies (a-Si:H, poly-Si, nc-Si/μc-Si:H), provided that the orientation allows the db to experience contact with a flexible part of the Si network as, e.g., present in regions with large H content or at surfaces and interfaces. Thereby, the trapping mechanism is accompanied by a bond-breaking process that, if stabilized, could act as the precursor state for charge-induced degradation in a-Si:H-based solar cells, a phenomenon that has been well known as Staebler-Wronski effect[76]).

The confirmation of a metastable recombination-precursor spin pair with weak spin-spin coupling, yet intrinsically strong localization, being a part of a charged db state in a quasi 2D-magic configuration, does not only resolve a long-standing debate about the nature of spin-dependent db recombination in silicon[23,42,77,78,79,80]. It also confirms the origin of the intriguing simultaneous existence of *strong localization and weak spin-spin coupling*. Similar to the recently reported coherent coupling of calcium monofluoride (CaF) molecules[99], it provides a kind of magic-angle driven entanglement of molecular spins, but here, thanks to the amorphous network, at much shorter distances and already at room temperature.

We find that the intrinsic properties of the electron spin pair constellation uncovered in the course of this study adds a well-known point defect in a technologically well-established materials class to the list of spin systems most promising for quantum information applications, in particular spin quantum sensing. Spin pairs based on the db$^{-*}$ state in the quasi 2D-magic configuration are straightforwardly initializable (electrically, by application of a recombination current), coherently controllable (by pulsed magnetic radiation fields) within long coherence times, and readable (electrically by measurement of a recombination current), while they are highly localized within a condensed matter network. Thus, the db state fulfills all requirements needed for spin-based sensing qubits in condensed matter, and, in contrast to other point defects whose suitability for quantum sensing has been established using optical excitation and manipulation, the db might be suitable for purely electrical single-spin readout. Moreover, it was shown on the nanoscopic level that db and bandtail states state play an active role in charge transfer processes at the a-Si:H/c-Si interface as was proven experimentally by conduction Atomic Force Microscopy (cAFM)[100]. This experiment provides evidence that single-defect contacting can be achieved, so that the here reported db state could open up new pathways towards silicon-based quantum-sensing applications at room temperature.

**Methods**

Sample preparation

Layered a-Si:H/μc-Si:H *pin* structures are deposited in a radio-frequency plasma-enhanced chemical vapor deposition (RF-PECVD) system at Forschungszentrum Jülich (Germany). The deposition procedure is described in detail in Ref. 81. Layers are deposited on 50 × 100 mm² Corning borosilicate glass substrates, covered with aluminum-doped ZnO transparent conductive oxide (TCO). The final structure consists of a 300 nm thick intrinsic a-Si:H layer, sandwiched between a *p*-type, boron-doped ($\approx 0.2$ at. %) and an *n*-type, phosphorus-doped ($\approx 0.2$ at. %) μc-Si:H injection layers with thicknesses of 20 and 30 nm, respectively. A final ZnO/Ag stack serves as electrical back contact.

To fit the samples into EDMR spectrometers, a special contact structure (see SI, Fig. S…a) has been developed and realized by laser scribing, as described in Refs. 28, 82 and 83. The sample area is confined to 1 × 1 mm², which is within the range of homogeneous microwave- and magnetic-field-amplitude distributions in the Bruker Flexline EPR resonator. Electrical contacts are provided by 100 nm thin and 50 mm long silver strips, minimizing electromagnetic field distortions within the microwave modes inside the resonator. Electrical connections to the detection system are established by wires attached contact strips using silver paste.

Electrically detected magnetic resonance

Multifrequency pulsed and continuous wave EDMR measurements are performed at EPR/EDMR spectrometers at University of Utah (100 MHz) and Helmholtz-Zentrum Berlin (9.6 GHz and 263 GHz).

EDMR at 100 MHz is measured on a home-built experimental setup (for details see Ref. 29). The sample is mounted on a sample rod and is positioned at the center of a 5-turn RF coil for homogeneous excitation. To sweep the external magnetic field ($B_0$), the current through a Helmholtz coil pair is controlled using a Kepco ATE 100-10M current source, controlled digitally by the digital-to-analog converter on a National Instruments PCI 6251 DAQ card. The device bias voltage is provided by an SRS SIM928 isolated voltage source and is maintained at a constant operating current of 500 μA. The device-current change at resonance is detected by a transimpedance amplifier (TIA) SRS570 with a 1 kHz–1 MHz bandpass filter and 20 μA/V gain. The output signal is acquired using an ATS9462 waveform digitizer. The excitation frequency of 100 MHz is generated by an Agilent MXG N5128A signal generator. The output of the source is connected to a 5 W Mini-Circuits ZHL-5W-1+ amplifier with 40 dB gain and a 5–500 MHz frequency range, delivering $\approx 79$ mW RF power at the position of the sample. The microwave source and the acquisition system are both triggered by a Pulse Blaster DDS-I-300 to orchestrate the overall timing. For the measurements shown in Fig. 1c, a microwave pulse of 4 μs duration was applied.

EDMR at 9.6 GHz is measured on a Bruker Elexsys E580 EPR spectrometer, upgraded with a home-built EDMR sample holder and current-detection system. The spectrometer is equipped with an X-band Bruker E580-1010 microwave bridge, a 1 kW travelling-wave-tube amplifier (Applied Systems Engineering 117X), a multi-channel pulse-forming unit (Bruker PatternJet), an 8-bit digitizer for data acquisition (Bruker SpecJet-II) and a sweepable 10″ iron magnet (Bruker ER 073), supplying $B_0$ fields of up to 1.45 T. The amplitude of $B_0$ is measured with an FT-NMR teslameter (Bruker ER 036TM). The sample is mounted and electrically contacted inside a dielectric-ring resonator (Bruker ER 4118X-MD-5) on a home-built sample holder.

For electrical detection, the contact strips of the sample are connected to a custom manufactured combined bias-voltage source and low-noise current-voltage converter and amplifier (Elektronik-Manufaktur Mahlsdorf, EMM). It contains a constant-voltage source to apply bias voltages ($V_B$) of up to ±3 V. The sample current is measured symmetrically between two voltage outputs $\pm V_B/2$, which suppresses interferences with electromagnetic background signals. Current-voltage conversion is achieved by a differential TIA with 100 µA/V gain. The absolute sample current is monitored with an SRS SIM970 digital voltmeter. The EDMR signal voltage is passed through a tunable bandpass filter (cutoff frequencies ranging from 1 Hz to 1 kHz and 200 kHz to 1.5 MHz, respectively) and further amplified by an adjustable gain factor of 5 to 2000, before the signal is fed into the digitizer. For the signal shown in Fig. 1d, a π-pulse with a length of 40 ns is used.

EDMR at 263 GHz is measured on a Bruker Elexsys E780 1016 setup, described in detail in Ref. 28. The spectrometer is equipped with a heterodyne microwave bridge, operating at 263 GHz (tunable by about ±1 GHz) and a cryogen-free superconductive magnet, providing $B_0$ fields between 0 and 12 T. The main coil of the magnet is stabilized at a constant magnetic field of 9.4 T. An additional sweep coil is used to conduct linear $B_0$ sweeps of ±150 mT. The microwave irradiation is coupled into a corrugated waveguide as a Gaussian beam via a quasi-optical front-end and propagated onto the sample that is placed in a non-resonant probehead. Electrical connection is established by thin gold wires attached by silver paste to the contact strips of the sample. The wires are soldered to a coaxial line, which is connected to the EMM detection system described above.

Due to the lack of a high-frequency microwave amplifier, the microwave power in the high-frequency setup is limited to 15 mW, resulting in maximum $B_1$ amplitudes of about 10 µT (depending on the alignment of the microwave beam) and resulting π-pulse lengths of 2–6 µs. For the spectrum shown in Fig. 1e, a pulse length of 4 µs is used.

Electrically detected Rabi nutations and Hahn-echo decay is measured at X-band using the above-described Bruker Elexsys E580 setup. Rabi nutations are recorded by integrating the transient current change after a single microwave pulse as a function of the pulse length[11,46,84,85]. The time-domain Rabi nutation of the integrated charge $\Delta Q$ is recorded as a function of the magnetic field $B_0$. Fast-Fourier transformation of the time-domain spectrum results in the magnetic-field/nutation-frequency maps shown in Fig. 2. Electrically detected Hahn echoes are recorded by applying a $\pi/2$–$\tau_1$–$\pi$–$\tau_2$–$\pi/2$ pulse sequence (see Fig. 1g), corresponding to the conventional EPR Hahn-echo sequence ($\pi/2$–$\tau$–$\pi$–$\tau$), supplemented by a final $\pi/2$ EDMR "detection pulse" that projects spin magnetization from the x-y-plane into z-direction (see Refs. 43–44 for details). A Hahn echo is measured by recording the integrated transient charge $\Delta Q$ as a function of $\tau_2$. To measure the Hahn-echo decay shown in Fig. 1g, the echo intensity is plotted as a function of the π-pulse delay $\tau_1$.

Least-squares fitting and numerical simulations

Spectral fits of multifrequency EDMR data (Figs. 1c–e) are obtained using MATLAB with the EasySpin toolbox[86]. We apply a global least-squares fitting approach to reduce the degrees of freedom for spin-Hamiltonian parameters that is described in detail in the SI (Note 1). Errors of the fitting parameters are estimated by a statistical bootstrap evaluation[87,88,89,90,91,92] (see SI, Note 2 for details). The Hahn-echo decay in Fig. 1g is fitted with a stretched exponential model $y = y_0 \exp[-(t/T_2)^\beta]$, where $\beta = 0.93(7)$. Errors of the fitting parameters $y_0$, $T_2$ and $\beta$ are likewise estimated by a bootstrap analysis.

Numerical simulations of ED-Rabi spectra are carried out in MATLAB by numerically solving the stochastic Liouville equation as described by Limes et al.[50] The simulation procedure and the modeling of different spin-pair formations is described in detail in Supplementary Notes 3 and 4.

DFT modeling

To model the db-related recombination process we apply density-functional theory (DFT) using the Quantum ESPRESSO package[93,94]. The a-Si:H network including dangling bond and tail states (cbt and vbt) are modeled using periodic boundary conditions using large supercells containing up to 1043 atoms, whereby the average H-content of the a-Si:H layer amounts to 18%. For modeling the a-Si:H network we start from the a-Si:H part of the interface structure of Ref. 26, which has been successfully shown to give a reasonable description, not only for db structures at the interface, but also for the cbt states in the a-Si:H network. As a result, the supercells comprise thin layers of a-Si:H with 1 nm thickness and a lateral expanse of up to 4.5 nm × 4.5 nm. Norm-conserving pseudo-potentials and a plane-wave basis set with 400 eV kinetic energy cutoff are used. Many-particle effects are taken into account with the Perdew-Burke-Ernzerhof (PBE) exchange-correlation functional[95]. For characteristic steps the EPR parameters are calculated using the GIPAW-module of the Quantum ESPRESSO package[94]. The elements of the electronic $g$ tensors are evaluated using a Berry-phase based method via the orbital magnetization[96], which allows to calculated the $g$ tensor also for the semi-localized states (e.g. bandtail states) or even metallic states. The $g$ tensors for the individual spins in the spin-pairs (Fig. 3 b-f) are calculated using constraint DFT, whereby the density of the second spin is distributed equally to both spin channels (and vice versa), so that only the spin under consideration contributes to the $g$ tensor. Since the ab initio calculation of the zero-field splitting ($D$ value) [72,73,101] is restricted to proper eigenstates of $S^2$, we are not able to give well-defined estimates of $D$ for the partially entangled state under consideration here. Hence, we focus on a theoretical evaluation of the exchange integral. For this purpose, besides ferromagnetically coupled high-spin (HS) configurations, also the antiferromagnetically coupled spin-pair configurations have to be modelled. Also known as broken-symmetry (BS) state, the total energy $E_{BS}$ of the latter is used to calculated the exchange coupling constant via[97] $J = (E_{HS} - E_{BS})/(\langle S^2 \rangle_{BS} - \langle S^2 \rangle_{HS})$.

**Acknowledgements**

We gratefully acknowledge Alexander Schnegg from MPI CEC in Mülheim for the stimulating discussions, the support with conducting the complex Rabi-nutation EDMR experiments at HZB. Without him these experiments would have never been conducted.

We would like to thank Friedhelm Finger and Oleksandr Asthakhov from Forschungszentrum Jülich (FZJ) as well as Jimmy Melskens and Arno Smets from Delft University of Technology for the preparation of EPR and EDMR samples. We also would like to thank Andreas von Kozierowski, Thomas Lußky and Martin Muske from HZB for their technical support.

U.G., T.B, W.G.S gratefully acknowledge Deutsche Forschungsgemeinschaft (DFG) from TRR 332 142/3-2024, Project No. 231447078 for funding as well as the Paderborn Center for Parallel Computing (PC2) for the provided computational resources.

H.P. was supported by the National Science Foundation, under award DMR #1701427. T.H.T. and H.M., were supported by the US Department of Energy, Office of Basic Energy Sciences, Division of Materials Sciences and Engineering under Award DE-SC0000909.

J.M., W.A., J.B., and K.L. gratefully acknowledge funding through the DFG priority programs SPP 1601 under contract numbers LI 2020/2-1 and LI 2020/2-2 as well as SPP 2314 under contract numbers LI 2020/3-1. We are also indebted to the Bundesministerium für Bildung und Forschung (BMBF) for their financial support through the network projects EPR-Solar (Grant No. 03SF0328) and EPRoC (Grant No. 03SF0565).

# Intrinsic silicon dangling bond spin-pair configuration for room temperature quantum sensing

J. Möser[1], H. Popli[2], T. H. Tennahewa[2], T. Biktagirov[3], J. Behrends[4], A. Schnegg[5], W. Akhtar[1], H. Malissa[2], C. Boehme[2], W.G. Schmidt[3], U. Gerstmann[3,*], K. Lips[1,4,*]

## *Supplementary Information*

**Figure S1.** Baseline correction of EDMR data

**Figure S2.** Bootstrap statistics for global multifrequency EDMR fitting

**Figure S3.** Simulation of ED-Rabi nutation patterns for different spin-pair models

**Figure S4.** Effect of $J$ and $D$ on the ED-Rabi nutation pattern (weak $B_1$)

**Figure S5.** Effect of $J$ and $D$ on the ED-Rabi nutation pattern (strong $B_1$)

**Figure S6.** Effect of $D$ anisotropy on the ED-Rabi nutation pattern

**Note 1.** Global least-squares fitting of multifrequency EDMR spectra

**Note 2.** Statistical bootstrap evaluation of least-squares fit parameters

**Note 3.** Numerical simulation of ED-Rabi nutation for spin-pair ensembles

**Note 4.** Simulation of ED-Rabi nutation patterns for different spin-pair models

**Table S1.** Spin-Hamiltonian parameters used for modeling different spin-pair ensembles.

[1]EPR4Energy & Berlin Joint EPR Lab, Department Spins in Energy Materials and Quantum Information Science, Helmholtz-Zentrum Berlin für Materialien und Energie, Hahn-Meitner-Platz 1, 14109 Berlin, Germany. [2]Department of Physics and Astronomy, University of Utah, Salt Lake City, Utah 84112, USA. [3]Lehrstuhl für Theoretische Physik, Universität Paderborn, Warburger Straße 100, 33098 Paderborn, Germany. [4]Berlin Joint EPR Lab, Fachbereich Physik, Freie Universität Berlin, Arnimallee 14, 14195 Berlin, Germany. [5]EPR4Energy, Max-Planck-Institut für chemische Energiekonversion, Stiftstraße 34-36, 45470 Mülheim an der Ruhr, Germany. *e-mail: lips@helmholtz-berlin.de; uwe.gerstmann@upd.de

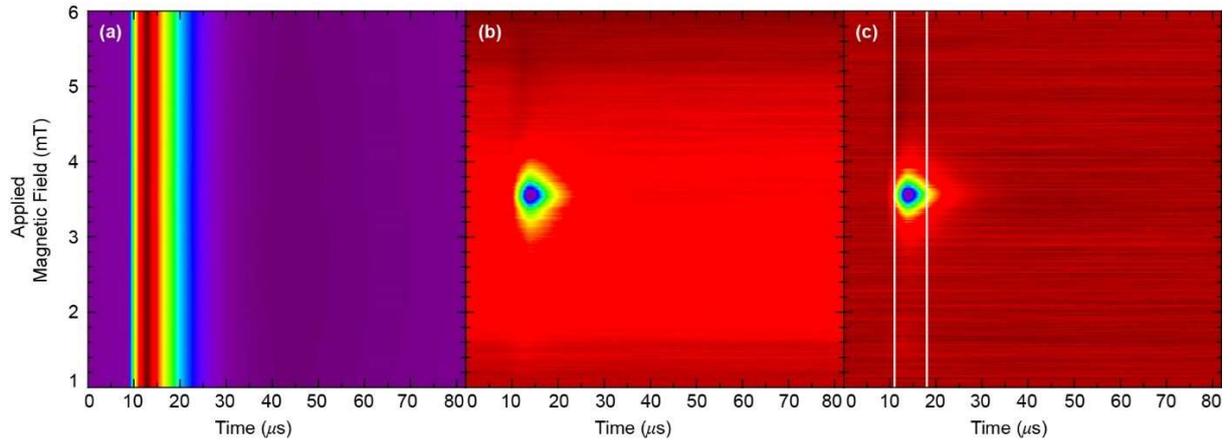

**Fig. S1. Baseline correction of EDMR data. (a)** EDMR transients of device-current changes at a microwave frequency of 100 MHz as a function of magnetic field. **(b)** The dataset from panel (a) after baseline correction. **(c)** Additional baseline correction obtained by subtraction of the final time slice, canceling any slow-varying background signal. The white lines indicate the numerical integration interval yielding the charge $Q$ that is plotted in Fig. 1c of the article.

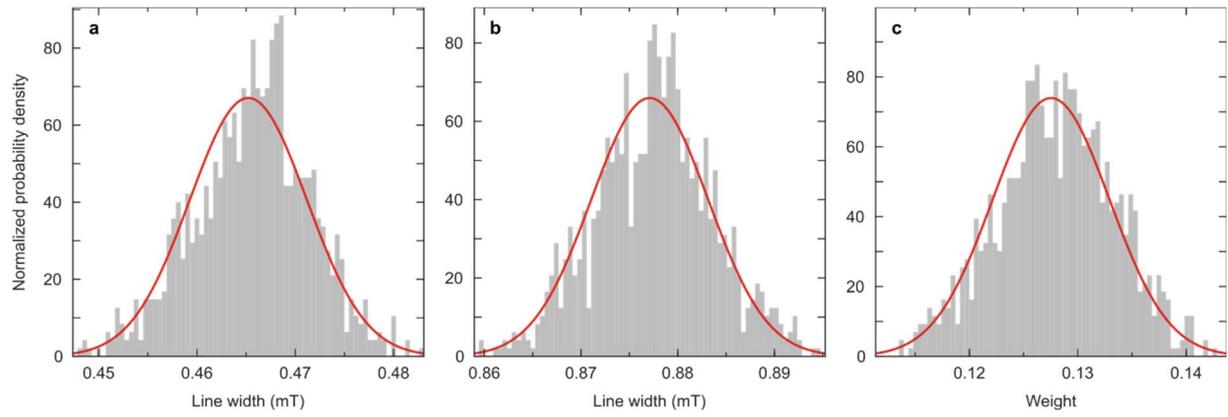

**Fig. S2. Bootstrap statistics for global multifrequency EDMR fitting.** Histograms for the global least-squares fit parameters **(a)** `lw_DB_lo`, the field-independent line width at low excitation frequency, **(b)** `lw_DB_hi`, the field-independent line width at high excitation frequency, and **(c)** `weight_VBT`, the weighting factor of the vbt component (see Supplementary Notes 1 and 2 for details). All distributions were obtained over *N* = 1000 bootstrap iterations. Red lines show the normal distribution with the arithmetic mean and standard deviation of the bootstrap distribution.

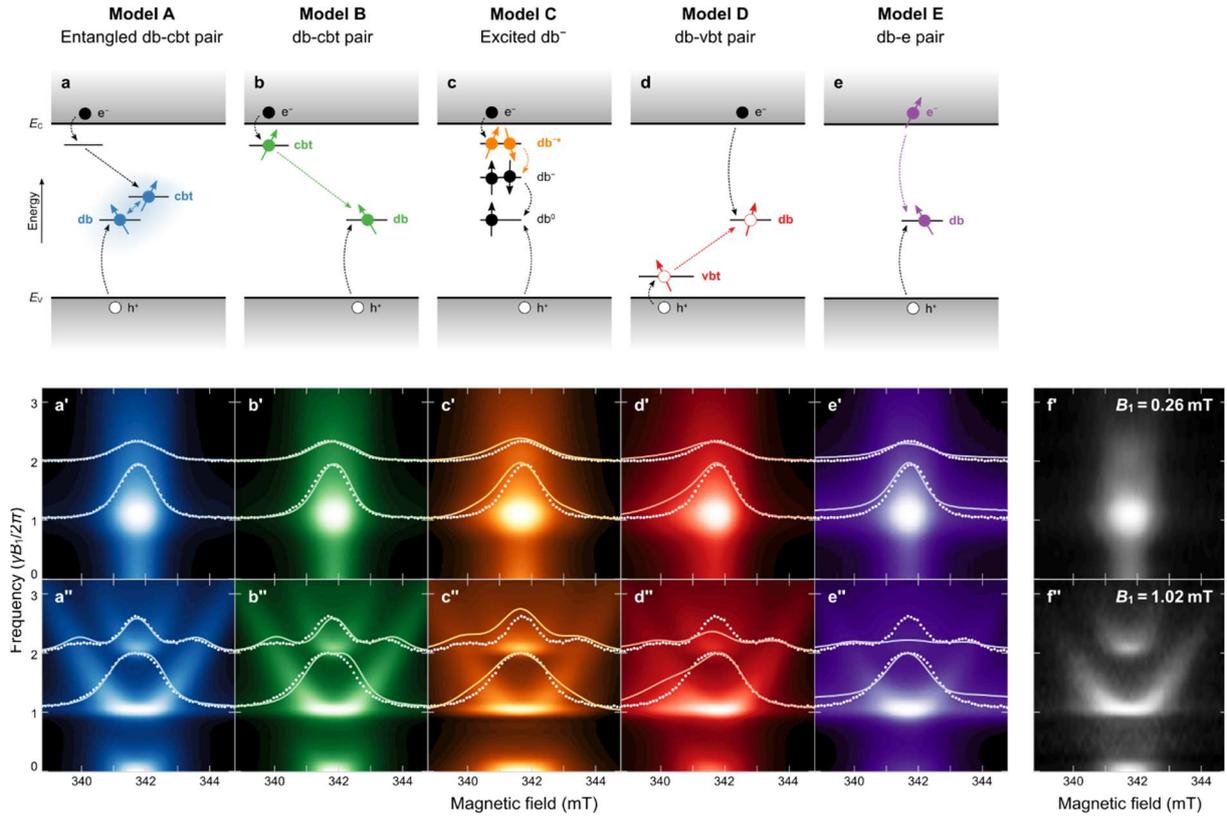

**Fig. S3. Simulation of ED-Rabi nutation patterns for different spin-pair models. Top: (a)–(e)** Schematic band-diagram representation of different models for spin-dependent db-assisted recombination in a-Si:H (see Note 4 for details). **Bottom:** Fast-Fourier transform (FFT) amplitudes of magnetic-field-resolved ED-Rabi nutations, numerically simulated for spin-pair distributions according to the models depicted above, and for microwave-excitation field strengths of **(a')–(e')** $B_1 = 0.26$ mT and **(a'')–(e'')** $B_1 = 1.02$ mT. **(f')** and **(f'')**: Measured magnetic-field/nutation-frequency map obtained as the FFT amplitude of magnetic-field-resolved ED-Rabi nutation experiments at the two respective $B_1$ excitation strengths. The one-dimensional traces in (a')–(e'') compare experimental (white dots) and simulated (colored lines) slices at Rabi frequencies $\Omega = \gamma B_1$ and $\Omega = 2\gamma B_1$.

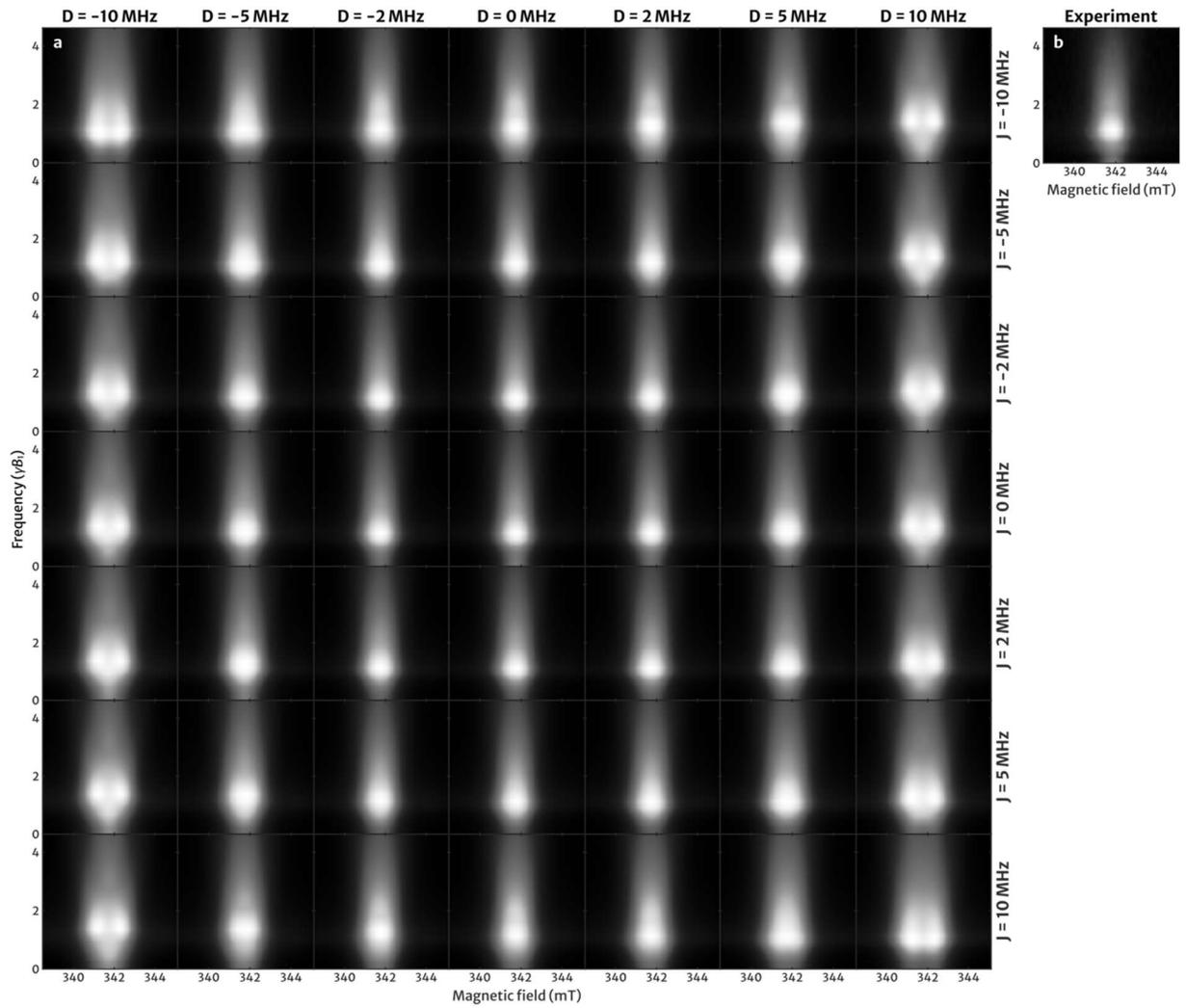

**Fig. S4. Effect of *J* and *D* on the ED-Rabi nutation pattern (weak $B_1$).** **(a)** Numerically simulated fast-Fourier transform (FFT) ED-Rabi magnetic-field/nutation-frequency maps for a spin-pair model comprising two db-like spins for different exchange- (*J*) and dipolar-coupling (*D*) strengths at microwave excitation $B_1 = 0.26$ mT. **(b)** Measured magnetic-field/nutation-frequency map obtained as the FFT amplitude of magnetic-field-resolved ED-Rabi nutation experiments at $B_1 = 0.26$ mT.

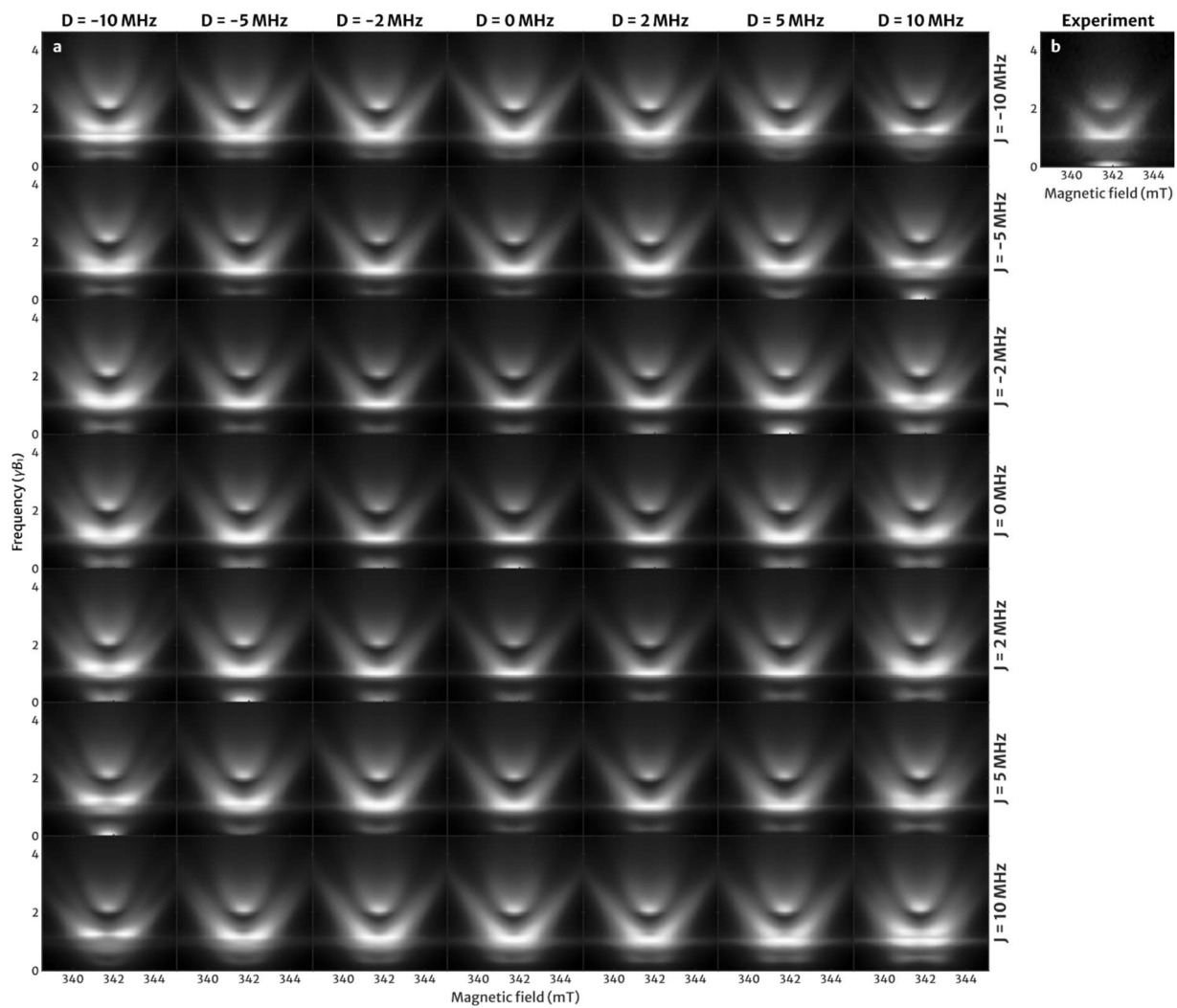

**Fig. S5. Effect of *J* and *D* on the ED-Rabi nutation pattern (strong $B_1$). (a)** Numerically simulated fast-Fourier transform (FFT) ED-Rabi magnetic-field/nutation-frequency maps for a spin-pair model comprising two db-like spins for different exchange- (*J*) and dipolar-coupling (*D*) strengths at microwave excitation $B_1$ = 1.02 mT. **(b)** Measured magnetic-field/nutation-frequency map obtained as the FFT amplitude of magnetic-field-resolved ED-Rabi nutation experiments at $B_1$ = 1.02 mT.

*Supplementary Note 1*

## Global least-squares fitting of multifrequency EDMR spectra

Line-shape modelling with EasySpin

*Accounting for magnetic-field offsets*

For global fitting models in particular, it is necessary to consider any shifts in magnetic field. Such shifts may arise from insufficient calibration of the applied magnetic field, from limited magnetic-field resolution, and from uncertainties in the *g*-value accuracy of the reference sample. Typically, the fit routine accounts for magnetic-field offsets by either of two ways: (i) The *g*-value is predefined and kept constant during the least-squares fit. Small shifts and offsets of the experimental spectra with respect to the magnetic field are compensated by allowing a field-offset fit parameter. (ii) The *g*-value is used as a fit parameter. In this case, one of the experimental spectra—typically, the one with the most accurate magnetic-field calibration—is used as the reference spectrum, for which the magnetic-field offset is set to zero. For all other experimental spectra, a non-zero field offset is allowed and included as a fit parameter.

The line-shape model function `pepper_fieldOffset`[S1], based on EasySpin's `pepper` function[S2], incorporates the above-described field-offset fitting procedure:

```
function varargout = pepper_fieldOffset(Sys, Exp, Opt)
    Exp1 = Exp;
    if isfield(Sys, 'fieldOffset')
        fieldOffset = Sys.fieldOffset;
        Sys1 = rmfield(Sys, 'fieldOffset');
        if isfield(Exp, 'Range')
            Exp1.Range = Exp.Range - fieldOffset;
        end
        if isfield(Exp, 'CenterSweep')
            Exp1.CenterSweep = Exp.CenterSweep;
            Exp1.CenterSweep(1) = Exp.CenterSweep(1) - fieldOffset;
        end
    else
        fieldOffset = 0;
        Sys1 = Sys;
    end
    [B, spec] = pepper(Sys1, Exp1, Opt);
    B1 = B + fieldOffset;
    switch nargout
        case 1
            varargout{1} = spec;
        case 2
            varargout{1} = B1;
            varargout{2} = spec;
    end
end
```

*Modelling the valence-band-tail spectrum*

The line-shape function associated with valence-band-tail (vbt) states (`pepper_VBT`) is based on `pepper_fieldOffset` and takes a field-offset parameter (`fieldOffset`), spectral weight (`weight_VBT`), *g*-tensor (`g_VBT`) and *g*-strain (`gStrain_VBT`) parameters, and a field-independent line width (`lw_VBT`) into account:

```
function varargout = pepper_VBT(Sys, Exp, Opt)
    Sys_VBT.fieldOffset = Sys.fieldOffset;
    Sys_VBT.weight = Sys.weight_VBT;
    Sys_VBT.g = Sys.g_VBT;
    Sys_VBT.gStrain = Sys.gStrain_VBT;
    Sys_VBT.lw = Sys.lw_VBT;
    [B, spec_VBT] = pepper_fieldOffset(Sys_VBT, Exp, Opt);
    switch nargout
        case 1
            varargout{1} = spec_VBT;
        case 2
            varargout{1} = B;
            varargout{2} = spec_VBT;
    end
end
```

*Modelling the dangling-bond spectrum*

The line-shape function for dangling-bond (db) defects (`pepper_DB`) takes a field-offset parameter (`fieldOffset`), *g*-tensor (`g_DB`) and *g*-strain (`gStrain_DB`) parameters, and a field-independent line width (`lw_DB`) into account. The total spectral weight is normalized to 100%, such that the weight of the db spectrum is given by `1-weight_VBT`. The low- and high-field line widths are given by the fit parameters `lw_DB_lo` and `lw_DB_hi`. The function `pepper_DB` selects the respective line width depending on the excitation frequency (with the cutoff being defined as 5 GHz, a somewhat arbitrary choice). In addition, the hyperfine splitting due to the presence of $^{29}$Si nuclei with a natural abundancy of 4.67 at. % is taken into account (`Nucs`, `AStrain`):

```
function varargout = pepper_DB(Sys, Exp, Opt)
    Sys_DB.fieldOffset = Sys.fieldOffset;
    Sys_DB.weight = 1 - Sys.weight_VBT;
    Sys_DB.g = Sys.g_DB;
    Sys_DB.gStrain = Sys.gStrain_DB;
    Sys_DB.Nucs = Sys.Nucs_DB;
    Sys_DB.A = Sys.A_DB;
    Sys_DB.AStrain = Sys.AStrain_DB;
    if Exp.mwFreq > 5
        Sys_DB.lw = Sys.lw_DB_hi;
    else
        Sys_DB.lw = Sys.lw_DB_lo;
    end
    [B, spec_DB] = pepper_fieldOffset(Sys_DB, Exp, Opt);
    switch nargout
        case 1
            varargout{1} = spec_DB;
        case 2
            varargout{1} = B;
            varargout{2} = spec_DB;
    end
end
```

*Modelling the combined dangling-bond & valence-band-tail spectrum*

The combined spectrum is a superposition of the db and vbt spectrum and is obtained by `pepper_DB_VBT`, which utilizes the above-defined `pepper_DB` and `pepper_VBT` functions:

```matlab
function varargout = pepper_DB_VBT(Sys, Exp, Opt)
    [B, spec_DB] = pepper_DB(Sys, Exp, Opt);
    [B, spec_VBT] = pepper_VBT(Sys, Exp, Opt);
    spec = spec_DB + spec_VBT;
    switch nargout
        case 1
            varargout{1} = spec;
        case 2
            varargout{1} = B;
            varargout{2} = spec;
    end
end
```

Global fitting with EasySpin

*Least-squares fitting routine for global models*

EasySpin[S2] includes a least-squares fitting routine (`esfit`), tailored to the modeling of EPR spectra. This routine is versatile and supports various fitting algorithms. In contrast to other fitting routines, `esfit` incorporates baseline correction and vertical scaling with different scaling functions, which reduces the number of fitting parameters and increases flexibility.

In this section, we describe `esfit_global`, which extends `esfit` for global modeling of multiple spectra simultaneously[S1,S3-S6]. This function accepts multiple datasets (`expSpc`) along with their respective experimental parameters (`Exp`) as well as the initial fit-parameter values (`Sys0`) and their respective bounds (`Vary`) as inputs. Other input parameters include the line-shape modeling function (`SimFnc`) as well as simulation (`SimOpt`) and fitting options (`FitOpt`). This least-squares algorithm allows for each experimental spectrum to have its own scaling function as well as local fit parameters that are independent for each dataset.

```matlab
function varargout = esfit_global(SimFnc, expSpc, Sys0, Vary, Exp, SimOpt, FitOpt)
    narginchk(5, 7);
    nargoutchk(0, 1);
    switch nargin
        case 5
            SimOpt1 = [];
            FitOpt1 = [];
        case 6
            SimOpt1 = SimOpt;
            FitOpt1 = [];
        case 7
            SimOpt1 = SimOpt;
            FitOpt1 = FitOpt;
    end
    SimOpt1.SimFnc = SimFnc;
    if numel(expSpc) ~= numel(Exp)
        error('expSpc and Exp must have the same number of components!');
    end
    SimOpt1.nExp = numel(Exp);
    SimOpt1.expSpc = expSpc;
    SimOpt1.Exp = Exp;
    if isfield(FitOpt1, 'Scaling')
        SimOpt1.Scaling = FitOpt1.Scaling;
    else
        SimOpt1.Scaling = 'none';
    end
    FitOpt1.Scaling = 'none';
    if isfield(FitOpt1, 'LocalParam')
        SimOpt1.LocalParam = FitOpt1.LocalParam;
        FitOpt1 = rmfield(FitOpt1, 'LocalParam');
    end
    flatSpc = cell2mat(expSpc);
    switch nargout
        case 0
            esfit('esfit_global_simfnc', flatSpc, Sys0, Vary, [], SimOpt1, FitOpt1);
        case 1
            BestSys = esfit(...
                'esfit_global_simfnc', flatSpc, Sys0, Vary, [], SimOpt1, FitOpt1);
            varargout{1} = BestSys;
    end
end
```

The helper function `esfit_global_simfnc` takes care of calling the modeling function (`SimFnc`) for each dataset with the respective set of fit parameters and concatenating the simulated spectra into a combined spectrum.

```matlab
    function [B, spec] = esfit_global_simfnc(Sys, ~, Opt)
        narginchk(3, 3);
        nargoutchk(1, 2);
        SimFnc = str2func(Opt.SimFnc);
        cellB = cell(Opt.nExp);
        cellSpec = cell(Opt.nExp);
        if isfield(Opt, 'LocalParam')
            LocalParam = Opt.LocalParam;
        end
        for i = 1:Opt.nExp
            Sys1 = Sys;
            Exp1 = Opt.Exp{i};
            if isfield(Opt, 'LocalParam')
                for j = 1:numel(LocalParam)
                    localParamName = LocalParam{j};
                    localParamArray = getfield(Sys1, localParamName);
                    localParamValue = localParamArray(i);
                    Sys1 = setfield(Sys1, localParamName, localParamValue);
                end
            end
            Opt1 = rmfield(Opt, {'SimFnc', 'nExp', 'expSpc', 'Scaling'});
            if isfield(Opt1, 'LocalParam')
                Opt1 = rmfield(Opt1, 'LocalParam');
            end
            [B1, spec1Unscaled] = SimFnc(Sys1, Exp1, Opt1);
            spec1 = rescale(spec1Unscaled, Opt.expSpc{i}, Opt.Scaling);
            if size(B1, 1) == 1
                cellB{i} = transpose(B1);
                cellSpec{i} = transpose(spec1);
            else
                cellB{i} = B1;
                cellSpec{i} = spec1;
            end
        end
        switch nargout
            case 1
                B = cell2mat(cellSpec);
            case 2
                B = cell2mat(cellB);
                spec = cell2mat(cellSpec);
        end
    end
```

Note that we had to use EasySpin version 5.2.27 due to changes to the `rescale` function in more recent versions.

5  *Global modeling of the experimental data*

The spin-system parameters for the db defects are based on previous results from multifrequency EPR reported by Fehr et al.[S7] and are kept fixed, with the exception of the line width (independently for low and high excitation frequencies, see above):

```matlab
    Sys0.g_DB = [2.0079, 2.0061, 2.0034];
    Sys0.gStrain_DB = [0.0054, 0.0022, 0.0018];
    Sys0.Nucs_DB = 'Si';
    Sys0.A_DB = [151, 151, 269];
    Sys0.AStrain_DB = [46, 46, 118];
    Sys0.lw_DB_lo = [0, 0.5];
    Vary.lw_DB_lo = [0, 0.5];
    Sys0.lw_DB_hi = [0, 1];
    Vary.lw_DB_hi = [0, 1];
```

10  The vbt states are modeled with:

```matlab
    Sys0.g_VBT = [2.014, 2.010, 2.006];
    Sys0.gStrain_VBT = [0.015, 0.007, 0.004];
    Sys0.lw_VBT = [0, 0.5];
    Sys0.weight_VBT = 0.15;
    Vary.weight_VBT = 0.15;
```

Since the *g*-values of both lines are kept fixed during the least-squares fit, it is necessary to include a field-offset parameter (see discussion above) for all three experimental spectra. These parameters are treated as local fit parameters specific to the respective spectrum.

```
Sys0.fieldOffset = [0, 0, 0];
Vary.fieldOffset = [0.1, 0.1, 1];
```

For the least-squares fit, we use the model function `pepper_DB_VBT` (see above). The experimental datasets are provided in `exp_spc_scaled`, which is a cell array of the measured spectra, roughly scaled to comparable amplitude, with approximately the same ratio of baseline to FWHM. The experimental parameters for each dataset are provided in `exp_param_scaled`, which is also a cell array.

```
FitOpt.LocalParam = {'fieldOffset'};
FitOpt.Method = 'simplex fcn';
FitOpt.Scaling = 'lsq1';
bestSys_global = esfit_global( ...
    'pepper_DB_VBT', ...
    exp_spc_scaled, ...
    Sys0, ...
    Vary, ...
    exp_param_scaled, ...
    [], ...
    FitOpt);
```

*Supplementary Note 2*

**Statistical bootstrap evaluation of least-squares fit parameters**

To verify the robustness of the global least-squares fit and determine confidence intervals for the fit parameters, we apply a bootstrap algorithm[S1,S3-S6,S8]. Bootstrap resampling is a widely used numerical approach that allows to derive statistical estimates for a single data sample. In essence, bootstrapping mimics a sample population by generating a synthetic distribution of $N$ samples through random resampling of the original data sample[8]. Statistics of least-squares fit parameters can then be estimated from the distribution of the respective bootstrap result.

We generate the bootstrap sample by resampling the residuals obtained from the best least-squares fit. Therefore, the simulated spectra resulting from the fit are subtracted from the experimental spectra to generate the residuals for each spectrum. Assuming that the deviations of the experimental from the simulated data points (i.e., the non-zero residual values) are random and independent from each other, the residuals constitute random noise only (n.b., without detailed knowledge of the nature of the noise). We then generate the set of $N$ artificial spectra by randomly resampling the residuals and adding the samples to the best-fit results. For each of these $N$ spectra, a global least-squares fit is performed, using the procedure described in Note 1. The resulting set of $N$ fit-parameter sets is used to estimate the statistical distribution and confidence intervals of the model parameters. The procedure is described in more detail in the SI of Ref. S1.

The MATLAB function call to run the bootstrap algorithm takes the form:

```
bootSys = bootstrap(...
    1000, ...
    'mixer_global', ...
    'esfit_global', ...
    'pepper_DB_VBT', ...
    simSpec_DB_VBT, ...
    residuals, ...
    bestSys_global, ...
    Vary, ...
    exp_param_scaled, ...
    [], ...
    FitOpt);
```

Here, `simSpec_DB_VBT` are the simulated spectra obtained from the parameter set `bestSys_global` and residuals are the differences between experimental and simulated spectra. The required MATLAB functions are:

```
    function bootSys = bootstrap(...
            nIter, ...
            MixFnc, ...
            FitFnc, ...
            SimFnc, ...
            simSpc, ...
            residuals, ...
            bestSys_global, ...
            Vary_global, ...
            Exp, ...
            SimOpt, ...
            FitOpt)
        mix_handle = str2func(MixFnc);
        fit_handle = str2func(FitFnc);
        bootSys = cell(1, nIter);
        for iIter = 1:nIter
            genSpec = mix_handle(simSpc, residuals);
            bootSys{iIter} = fit_handle(...
                SimFnc, genSpec, bestSys_global, Vary_global, Exp, SimOpt, FitOpt);
        end
    end
```

```
    function genSpec = mixer_global(simSpec, residuals)
        genSpec = cell(size(residuals));
        for i = 1:length(residuals)
            genSpec{i} = mixer(simSpec{i}, residuals{i});
        end
    end
```

```
    function genSpec = mixer(simSpec, residuals)
        genSpec = zeros(size(residuals));
        p = randi(length(residuals), size(residuals));
        for i = 1:length(residuals)
            genSpec(i) = residuals(p(i)) + simSpec(i);
        end
    end
```

The bootstrap distributions for $N = 1000$ bootstrap samples of the three main fit parameters, the two line-width parameters `lw_DB_lo` and `lw_DB_hi` as well as the vbt weighting factor `weight_VBT` are shown in *Fig. S2*. The bootstrap distributions for all fit parameters are normally distributed (red lines in *Fig. S2*), exhibiting no significant sign of bias or asymmetry. The standard deviations of the normal distributions thus serve as estimates of the 68% bootstrap confidence intervals, i.e., the uncertainty in the least-squares fit-parameter results. For the two line-width parameters `lw_DB_lo` and `lw_DB_hi`, we obtain values of 0.465(7) mT and 0.877(7) mT, respectively; the resulting weight parameter `weight_VBT` is 0.127(6).

*Supplementary Note 3*

# Numerical simulation of ED-Rabi nutations for spin-pair ensembles

Single-pair solution of the stochastic Liouville equation

Electrically detected Rabi nutations for a single $S = 1/2$ pair are simulated by numerically solving the stochastic Liouville equation, applying the superoperator formalism introduced by Limes et al.[9].

In MATLAB, the spin pair is defined by a structure Sys, where Sys.g is a two-element vector containing the *g*-values of both spins, Sys.J is the exchange coupling (in megahertz), and Sys.D and Sys.E are the dipolar-coupling parameters *D* and *E*. In addition, the experimental parameters are given in a structure Exp, defining the microwave frequency (Exp.mwFreq, in gigahertz), the microwave field amplitude $B_1$ (Exp.B1, in millitesla) and the external magnetic field $B_0$ (Exp.B0, in millitesla). The field-independent terms of the spin-pair Hamiltonians, $H_0$ and $H_1$, are then calculated as follows:

```matlab
MT_TO_MHZ = 7.1448e-20;

% Spin-operator matrices (uses EasySpin!)
[Sx1, Sy1, Sz1, Sx2, Sy2, Sz2] = sop(Sys.S, 'x1', 'y1', 'z1', 'x2', 'y2', 'z2');
Sx12 = Sx1 * Sx2;
Sy12 = Sy1 * Sy2;
Sz12 = Sz1 * Sz2;

% Dipolar-coupling vector
Sys.eeD = Sys.D*[-1 -1 2] + Sys.E*[1 -1 0];

% Field-independent spin-Hamiltonian terms
H0_coupling = Sys.J*S12 + Sys.eeD(1)*Sx12 + Sys.eeD(2)*Sy12 + Sys.eeD(3)*Sz12; % MHz
H0_zeeman   = MT_TO_MHZ * (Sys.g(1)*Sz1 + Sys.g(2)*Sz2);                      % MHz/mT
H0_rot      = -Exp.mwFreq*1e3 * Sz;                                            % MHz
H1          = Exp.B1*MT_TO_MHZ * (Sys.g(1)*Sx1 + Sys.g(2)*Sx2);                % MHz
```

In addition, we calculate the high- (triplet, $S = 1$) and low-spin (singlet, $S = 0$) subspace operators using the helper function coupledSpinOperators(), which takes a two-element vector s holding the spin quantum numbers of the two coupled spins (s(1) = s(1) = 1/2 for the present case of spin-1/2 pair).

```matlab
U2C = coupledSpinOperators(Sys.S);
Ut = U2C{1}; % triplet subspace operator
Us = U2C{2}; % singlet subspace operator
```

```matlab
function U2C = coupledSpinOperator(s)

    % Clebsch-Gordan coefficient matrix (uses EasySpin!)
    C = cgmatrix(s(1), s(2));

    % Coupled spin states
    S_vals = abs([sum(s); diff(s)]);
    N = 2*S_vals + 1;
    S = cell2mat(arrayfun(@(s, n) repmat(s, n, 1), S_vals, N, 'UniformOutput', false));
    Ms = cell2mat(arrayfun(@(s) (s:-1:-s).', S_vals, 'UniformOutput', false));

    % Populate high- (U2C{1}) and low-spin (U2C{2}) operator matrices
    U2C = repmat({0}, 1, 2);
    for i = 1:sum(n)
        psi = zeros(sum(n), 1);
        psi(i) = 1;
        k = S_vals == S(i);
        U{k} = U{k} + bsxfun(@times, C*psi, psi'*C);
    end
end
```

The time-domain signal matrix is then populated by numerically solving the stochastic Liouville equation at each magnetic-field point.

```
z = zeros(nTime, nField);
for iField = 1:nField
    H0 = Exp.B0(iField) * H0_zeeman + H0_coupling + H0_rot; % full static spin Hamiltonian (MHz)
    z(:, iField) = rabiSuper(H0, H1, Ut, Us, Sys, Exp);
end
```

The function `rabiSuper()` is based on the superoperator formalism introduced by Limes et al.[S9], which is described in detail in Appendix A of Ref. S10.

```
function z = rabiSuper(H0, H1, Ut, Us, Sys, Exp)

    % Transform Hamiltonian and spin operators into the energy eigenbasis (uses EasySpin!)
    [U, E] = eig(H0);
    U_ = U';
    H = U_ * (H0 + H1) * U;
    Ut = U_ * Ut * U;
    Us = U_ * Us * U;

    % Calculate the observable dQ. Here, gamma is the total spin-pair annihilation rate
    % (dissociation + recombination) and prob is the relative dissociation probability
    % d_i/(d_i + r_i).
    d = Sys.Dissociation(1)*Ut + Sys.Dissocation(2)*Us;
    r = Sys.Recombination(1)*Ut + Sys.Recombination(2)*Us;
    gamma = diag(d + r);
    prob = diag(d) ./ gamma;
    dQ = diag(prob .* exp(-gamma*Exp.tau(1)) - exp(-gamma*Exp.tau(2))));

    % Create the initial density matrix.
    if ~isempty(Exp.Temperature)
        % Optional Boltzmann-polarization weights
        w = exp(-E / (Exp.Temperature * boltzm/planck*1e-6));
        w = w / sum(w);
    else
        w = 1/4; % Equally weighted spin-pair generation
    end
    rho0 = diag(Sys.Generation * w ./ gamma)
    rho0 = rho0 / sum(rho0(:));

    % Superoperator and super density matrix
    H = kron(expm(1i*2*pi*Exp.dt * H), expm(-1i*2*pi*Exp.dt * H.'));
    rho0 = reshape(rho, [], 1);
    rho = rho0;
    dQ = reshape(dQ, [], 1)';

    % Time evolution in steps of dt
    n = length(Exp.Time);
    z = zeros(n, 1);
    for i = 2:n
        rho = H * rho;
        z(i) = real(dQ * (rho - rho0));
    end
end
```

Modeling of spin-pair ensembles

The above-described code numerically simulates time-domain field-swept Rabi nutations for a single pair of spins `Sys.S` with $g$-values `Sys.g` and spin-spin coupling `Sys.J` and `Sys.D`/`Sys.E`. In real systems, a distribution of spin pairs is formed from distributions of spins (e.g., db and cbt spins forming a distribution of db-cbt pairs in a-Si:H). We model such spin-pair distributions by calculating discrete weighted $g$-value distributions for the two paired spins, based on their known EPR spectra, and form spin pairs as combinations of the two spin distributions.

For instance, the distribution of db spins can be estimated from the spin Hamiltonian parameters known from multifrequency EPR[S7].

```
    Sys.S = 0.5;
    Sys.g = [2.0079 2.0061 2.0034];
    Sys.gStrain = [0.0054 0.0022 0.0018];
    Sys.Nucs = 'Si';
    Sys.A = [151 269];
    Sys.AStrain = [46 118];
    Sys.lw = [0.13 0.43];
```

Using EasySpin's `pepper()` function, we simulate the field-domain EPR spectrum at the given (X-band) microwave frequency and use the spectrum as a *g*-value distribution of db spins, where the spectral intensity serves as the weight of the given *g*-value.

```
    [B0, w0] = pepper(Sys, Exp);
    g0 = planck * Exp.mwFreq ./ (bmagn * B0) * 1e12;

    % Interpolation to equally spaced g-axis
    g = linspace(min(g0), max(g0), Exp.nPoints);
    w = interp1(g0, w0, g, 'pchip');
```

Combining two *g*-value distribution to a spin-pair ensemble, the parameter space—and thus computation time—grows quadratically with the number of *g*-values. To decrease the number of discrete spin-pair *g*-value combinations to a manageable amount, we slice the above distribution `g` into n *equally weighted* slices. (Alternatively, one could slice the g-value distribution into *equally spaced* slices with the according weighting factors. As a result, however, a large part of the simulations for *g*-values stemming from the spectral edges would only contribute at diminishing weights, such that equally weighted slices are computationally more efficient.)

```
    [g, w] = buildEqualWeightSlices(g0, w0, n);
```

```
    function [g, w] = buildEqualWeightSlices(g0, w0, n)
        % Split the distribution into n slices with weights dw = 1/n. Therefor, w0 is numerically
        % integrated and then stepwise interpolated to multiples of dw. The g-value of each slice
        % is found by calculating the weighted arithmetic mean.
        w_int = abs(cumtrapz(w0));
        w_int = w_int / max(w_int);

        % Limit range to 0 < w_int < 1.
        k = find(w_int <= 0.0, 1, 'last') : find(w_int >= 1.0, 1, 'first');
        g0 = g0(k);
        w0 = w0(k);
        w_int = w_int(k);

        % Build the slices.
        dw = 1 / n;
        g = zeros(1, n);
        w = dw * ones(1, n);
        g1 = g0(1); % lower edge of the current slice
        for i = 1:n
            g2 = interp1(w_int, g0, i*dw);      % upper edge of the current slice
            gi = linspace(g1, g2, length(g0));  % interpolation axis within the current slice
            wi = interp1(g0, w0, gi, 'pchip');  % interpolated weights within the current slice
            g(i) = sum(gi .* wi) / sum(wi);     % weighted arithmetic mean of the slice's g-values
            g1 = g2;
        end
    end
```

Given two discrete g-value distribution for the two paired spins, the g-value parameter space of the pair ensemble is simply the set of all combinations of the two spin ensembles (assuming no correlation between the two distributions).

```
    g = allcomb(g1, g2);
    w = prod(allcomb(w1, w2), 2);
```

Here, `allcomb()` is a helper function published by J. van der Geest at MATLAB Central File Exchange[S11]. Note that the calculation of the weights `w` is unnecessary for equally weighted slices. However, in the original code we allowed for both equally weighted and equally spaced slice to exclude any numerical artifacts stemming from the method of building the parameter space.

To incorporate the *g*-value parameter space into the simulation code, the *g*-dependent code block is wrapped into another loop (`parfor`, using the MATLAB Parallel Computing Toolbox) and the total ensemble simulation is summed with the appropriate weighting factors.

```matlab
% Field-independent spin-Hamiltonian terms
H0_coupling = ...
H0_rot = ...
z = 0;
parfor i = 1:length(g)
    gi = g(i, :);
    wi = w(i, :);
    
    % Simulation for a single spin pair at weight wi.
    H0_zeeman = MT_TO_MHZ * (Sys.gi(1)*Sz1 + Sys.gi(2)*Sz2); % MHz/mT
    H1 = Exp.B1*MT_TO_MHZ * (Sys.gi(1)*Sx1 + Sys.gi(2)*Sx2;
    zi = zeros(nTime, nField);
    for iField = 1:nField
        H0 = Exp.B0(iField) * H0_zeeman + H0_coupling + H0_rot; % MHz
        zi(:, iField) = rabiSuper(H0, H1, Ut, Us, Sys, Exp);
    end
    
    % Add to the total result at the given weight.
    z = z + wi*zi;
end
```

*Supplementary Note 4*

**Simulation of ED-Rabi nutation patterns for different spin-pair models**

*Figure S3* shows simulated ED-Rabi nutation pattern that are based on different models for the observed spin-dependent db-related recombination mechanism. The fundamental difference between the models is the formation of the spin pair between different localized states is a-Si:H. This leads to different spin-pair *g*-value ensembles forming the parameter space for the simulation (*see Note 3*).

Model A in *Figs. S3a–a″* corresponds to the assigned model discussed in the manuscript. Based on the EDMR spectrum being predominantly formed by a single db resonance and the DFT-predicted g-value adaption of the cbt state, the spin-pair ensemble is modeled as two (uncorrelated) db ensembles. The db *g*-value distribution is calculated based on the spin-Hamiltonian parameters known from multifrequency EPR[S7] (see *Fig. S1*). The two db ensembles are then combined to a spin-pair ensemble as discussed above (*Note 3*).

In model B (*Figs. S3b–b″*), a spin pair is formed between a neutral db and an electron trapped in a nearby shallow cbt state that does *not* adapt the db *g*-tensor, but maintains its g-value ($g_{cbt}$ = 2.0044 as known from light-induced EPR, LEPR[S12]). Such a cbt-db pair has been suggested to drive the spin-dependent transition by Dersch and others[S13,S14]. In this case, the spin-pair ensemble is formed between a distribution of db spins[S7] and a distribution of cbt spins with an isotropic $g_{cbt}$ = 2.0044 (see *Tab. S1*). The experimentally observed relative intensities of the two Rabi features at $\Omega = \gamma B_1$ and $\Omega = 2\gamma B_1$ is reproduced by both model A and B. The simulated $\Omega = 2\gamma B_1$ intensity in *Figs. S3b′* and *b″* is only marginally smaller than in *Fig. S3a″*, as the average *g*-values $g_{cbt}$ = 2.0044 and $g_{db}$ = 2.0055 are close such that the average *g*-value difference between the paired spins is similar for the two spin-pair ensembles in models A and B. However, reveal that the cbt-db model produces a field offset between experiment and simulation due to the contribution of cbt spins. This observation agrees with the absence of a cbt contribution in the multifrequency pEDMR spectra (*Fig. 1* in the article) and contradicts the cbt-db pair model.

An alternative model (model C, *Figs. S3c–c″*), relying on the formation of an *excited* db⁻ state has been proposed for db-assisted recombination in µc-Si:H films and at the SiO$_2$/c-Si interface[14-16]. It assumes direct capture of a band electron by a neutral db, forming an excited doubly occupied db⁻. The spin dependency is not embedded in this capture process, but in the *readjustment time* that the excited state takes to relax into the db⁻ ground state, under the constraint of spin-selection rules. To our knowledge, it remains unclear how the *g*-tensor of the excited db⁻ state would look like. However, with the paired spins occupying the same electronic state, we assume a strong correlation between their *g*-tensors. Assuming the db spin Hamiltonian (*Tab. S1*), we model this correlation by constraining the Larmor-frequency separation between the paired spins by a Lorentzian line profile with a FWHM of 0.5 mT, corresponding to the approximate natural line width of the db resonance known from EPR[7]. The change of the ensemble's average *g*-value separation by the introduced correlation strongly influences the spectral shape and intensity of the $\Omega = \gamma B_1$ and $\Omega = 2\gamma B_1$ peaks in the

|  | g-tensor | | | $^{29}$Si HFI-tensor A (MHz) | | | Line width (mT) | | Ref. |
|---|---|---|---|---|---|---|---|---|---|
|  | $g_x$ or $g_\perp$ [strain] | $g_y$ or $g_\perp$ [strain] | $g_z$ or $g_\parallel$ [strain] | $A_x$ or $A_\perp$ [strain] | $A_y$ or $A_\perp$ [strain] | $A_z$ or $A_\parallel$ [strain] | $\Delta B_G$ | $\Delta B_L$ |  |
| db$^0$ | 2.0079 [0.0054] | 2.0061 [0.0022] | 2.0034 [0.0018] | 151 [46] | 151 [46] | 269 [118] |  | 0.47 | Fehr[7] |
| cbt | 2.0044 [0.0033] | 2.0044 [0.0033] | 2.0044 [0.0033] | 200 [150] | 200 [150] | 200 [150] |  | 0.5 | Umeda[11] |
| vbt | 2.0171 [0.0136] | 2.0091 [0.0079] | 2.0041 [0.0030] |  |  |  |  | 0.64 | Umeda[11] |
| e$^-$ | 2.0044 | 2.0044 | 2.0044 |  |  |  | 10 |  |  |

**Tab. S1. Spin-Hamiltonian parameters used for modeling different spin-pair ensembles.** See text for details.

simulated Rabi spectra (*Figs. S3c′* and *c″*). Both peaks are strongly broadened with respect to the experimental data. In addition, the relative intensity of the $\Omega = 2\gamma B_1$ (spin-locking) peak clearly increases. This is in agreement with the findings by

Michel et al. for disordered spin-pair ensembles with correlated g-values[S18]. Besides the g-value correlation that results in the evident discrepancy between the experimental and the numerically simulated Rabi spectra in *Figs. S3c′* and *c″*, the assumption of both spins occupying a single excited db$^-$ state suggests dipolar and exchange coupling being present between the paired spins. While all simulations shown in *Fig. S3* assume $D = J = 0$, we have demonstrated in *Fig. S2* (main article) and *Figs. S4* and *S5* that the introduction of $D$ and $J$ coupling produces features in the Rabi spectra that are clearly not observed experimentally. Therefore, we can rule out the direct-capture model into an excited db$^-$ state to explain the observed spin-dependent current in a-Si:H.

In model D (*Figs. S3d–d″*), we assume the spin pair being formed between the db$^0$ and a hole trapped in a nearby vbt state. After the spin-dependent transition, the recombination is completed by the capture of a conduction-band electron into the unoccupied db$^+$ state (*Fig. S3d*). For such a model the spin-pair ensemble is formed between db$^0$ spins, using the spin-Hamiltonian parameters from *Tab. S1*, and the ensemble of vbt spins. The g-tensor of vbt states is highly anisotropic with an average $g_{vbt} \approx 2.01$, as is known from LEPR[11]. However, to our knowledge, precise values for the g-tensor's principal values are not available in the literature. We thus digitized multifrequency LEPR data measured at 3-34 GHz by Umeda et al. (Ref. S12, Fig. 1) and fitted the spectra with a global least-squares routine, as described in *Note 1*, using a spin-Hamiltonian model comprising a rhombic g-tensor, g-strain and a field-independent Lorentzian line width. (We neglect HFI as the HFI has found to be weak for vbt states[S12,S19].) The results of this fit are listed in *Tab. S1* and will be published in more detail in a subsequent publication. The simulated Rabi-nutation spectra for such an ensemble of db-vbt pairs are shown in *Fig. S3d′–d″*. Due to the larger average g-value separation between db and vbt spins, the beating component at $\Omega = 2\gamma B_1$ is much less pronounced than in models A–C and as experimentally observed. In the addition, the spectral shapes that emerge due to the vbt resonance, significantly deviate from the experimental Rabi-nutation spectra, such that spin-dependent db-vbt transitions can be certainly excluded as the principal origin of the EDMR signal.

Finally, we have included a model (model E, *Figs. S3e–e″*), that explain the absence of a second (non-db) resonance with a broadened g-value distribution of the paired electron spin, e.g., due to delocalization and fast spin relaxation, such that the resonance becomes less apparent in the EDMR spectrum. We modeled this by assuming the isotropic $g_{cbt} = 2.0044$ and a simple

**20**

phenomenological Gaussian line width of 10 mT. The simulated Rabi spectra for the resulting spin-pair ensemble is shown in *Figs. S3e′–e″*. It becomes immediately apparent that this model cannot explain the observed signal. The broad *g*-value distribution of the e⁻ spins strongly increases the average *g*-value separation of the paired spins. As a result, the spin-locking component ($\Omega = 2\gamma B_1$) almost vanishes, even at high $B_1$ strength (*Fig. S3e″*), clearly conflicting with the experimental data.

The above evaluation of different spin-pair models substantiates our finding that the spectral shape as well as the intensities of the observed Rabi-nutation spectra can only be explained by a weakly coupled spin-pair model, were both spins exhibit the EPR signature of dbs.